\documentclass{article}[11pt]
\usepackage{jheppub}
\usepackage[toc,page]{appendix}
\pdfoutput=1
\usepackage{graphicx}
\usepackage{amsmath,amssymb,mathrsfs}
\usepackage{bbm}
\usepackage{color}
\usepackage{xcolor}
\usepackage{dsfont}
\usepackage{cancel}
\usepackage{dsfont}
\usepackage{epstopdf}
\usepackage{epsfig}
\usepackage{bm}
\usepackage{dcolumn}
\usepackage{hyperref}
\usepackage{enumitem}
\usepackage{multirow}
\usepackage{lineno}
\usepackage{mathtools,slashed}
\usepackage{url}
\usepackage{lineno}
\usepackage{wrapfig}

\newcommand{\ben}{\begin{enumerate}}
\newcommand{\een}{\end{enumerate}}
\newcommand{\bit}{\begin{itemize}}
\newcommand{\eit}{\end{itemize}}

\newcommand{\beqa}{\begin{eqnarray}}
\newcommand{\eeqa}{\end{eqnarray}}
\newcommand{\beq}{\begin{equation}}
\newcommand{\eeq}{\end{equation}}
\newcommand{\bay}{\begin{array}}
\newcommand{\eay}{\end{array}}

\def\ifmath#1{\relax\ifmmode #1\else $#1$\fi}

\def\gsim{\ \rlap{\raise 3pt \hbox{$>$}}{\lower 3pt \hbox{$\sim$}}\ }
\def\lsim{\ \rlap{\raise 3pt \hbox{$<$}}{\lower 3pt \hbox{$\sim$}}\ }

\def\ls#1{\ifmath{_{\lower1.5pt\hbox{$\scriptstyle #1$}}}}
\def\lsup#1{^{\lower 6pt\hbox{$\scriptstyle#1$}}}

\def\bracket#1#2 {\mathinner{\langle{#1}|{#2}\rangle}}

\def\bracket#1#2 {\mathinner{\langle{#1}|{#2}\rangle}}

\newcommand{\mysubsubsection}[1]{{\it #1~}}

\newcommand{\be}{\begin{equation}}
\newcommand{\ee}{\end{equation}}
\newcommand{\bea}{\begin{eqnarray}}
\newcommand{\eea}{\end{eqnarray}}

\graphicspath{{figs/}}
\begin{document}

%\linenumbers

\title{AEDGE:\\
Atomic Experiment for Dark Matter and Gravity Exploration in Space}

\abstract{
We propose in this White Paper a concept for a space  experiment using cold atoms to search for ultra-light
dark matter, and to detect gravitational waves in the frequency range between the most sensitive
ranges of LISA and the terrestrial LIGO/Virgo/KAGRA/INDIGO experiments. This interdisciplinary experiment, called
Atomic Experiment for Dark Matter and Gravity Exploration (AEDGE), will also complement other planned
searches for dark matter, and 
%will have many 
 exploit synergies with other gravitational wave detectors. We give examples of the extended range of sensitivity to ultra-light dark matter offered by AEDGE, and how its gravitational-wave measurements could explore the assembly of super-massive black holes, first-order phase transitions in the early universe and cosmic strings.
AEDGE will be based upon technologies now being developed for terrestrial experiments using
cold atoms, and will benefit from the space experience obtained with, e.g., LISA and cold atom experiments in microgravity.\\
~~\\
~~\\
~~\\
KCL-PH-TH/2019-65, CERN-TH-2019-126
}

%\affiliation[*]{White Paper Editor}
\affiliation[@]{White Paper Contact Person}
\affiliation[*]{White Paper Author}
%%%%%%%%%%%%%%%%%%%%%
\author[1]{Yousef~Abou~El-Neaj,}
%%\email{andrea.bertoldi@institutoptique.fr}
\affiliation[1]{Physics Department, Harvard University, Cambridge, MA 02138, USA}
%%%%%%%%%%%%%%%%%%%%%%

%%%%%%%%%%%%%%%%%%%%%
\author[2]{Cristiano~Alpigiani,}
%%\email{andrea.bertoldi@institutoptique.fr}
\affiliation[2]{Department of Physics, University of Washington, Seattle, WA 98195-1560, USA}
%%%%%%%%%%%%%%%%%%%%%%

%%%%%%%%%%%%%%%%%%%%%
\author[3]{Sana~Amairi-Pyka,}
%%\email{andrea.bertoldi@institutoptique.fr}
\affiliation[3]{Humboldt Universit{\" a}t zu Berlin, Institute of Physics, Newtonstrasse 15, 12489 Berlin, Germany}
%%%%%%%%%%%%%%%%%%%%%%

%%%%%%%%%%%%%%%%%%%%%
\author[4]{Henrique~Ara\'{u}jo,}
%%\email{andrea.bertoldi@institutoptique.fr}
\affiliation[4]{Blackett Laboratory, Imperial College London, Prince Consort Road, London, SW7 2AZ, UK}
%%%%%%%%%%%%%%%%%%%%%%

%%%%%%%%%%%%%%%%%%%%%
\author[5]{Antun~Bala{\v z},}
%%\email{andrea.bertoldi@institutoptique.fr}
\affiliation[5]{Institute of Physics Belgrade, University of Belgrade, Pregrevica 118, 11080 Belgrade, Serbia}
%%%%%%%%%%%%%%%%%%%%%%

%%%%%%%%%%%%%%%%%%%%%
\author[6]{Angelo~Bassi,}
%%\email{andrea.bertoldi@institutoptique.fr}
\affiliation[6]{Dipartimento di Fisica, Universit{\` a} degli Studi di Trieste, I-34127 Trieste, Italy}
%%%%%%%%%%%%%%%%%%%%%%

%%%%%%%%%%%%%%%%%%%%%
\author[7]{Lars~Bathe-Peters,}
%%\email{andrea.bertoldi@institutoptique.fr}
\affiliation[7]{Institut f{\" u}r Theoretische Physik, Technische  Universit{\" a}t  Berlin, Hardenbergstraße 36, 10623 Berlin, Germany}
%%%%%%%%%%%%%%%%%%%%%%

%%%%%%%%%%%%%%%%%%%%%
\author[8]{Baptiste~Battelier,}
%%\email{andrea.bertoldi@institutoptique.fr}
\affiliation[8]{LP2N, Laboratoire Photonique, Num\'erique et Nanosciences, Universit\'e Bordeaux-IOGS-CNRS:UMR 5298, F-33400 Talence, France}
%%%%%%%%%%%%%%%%%%%%%%

%%%%%%%%%%%%%%%%%%%%%
\author[5]{Aleksandar~Beli\'{c},}
%%\email{andrea.bertoldi@institutoptique.fr}
%\affiliation[5]{Institute of Physics Belgrade, University of Belgrade, Pregrevica 118, 11080 Belgrade, Serbia}
%%%%%%%%%%%%%%%%%%%%%%

%%%%%%%%%%%%%%%%%%%%%
\author[9]{Elliot~Bentine,}
%%\email{andrea.bertoldi@institutoptique.fr}
\affiliation[9]{Clarendon Laboratory, University of Oxford, Oxford OX1 3RH, UK}
%%%%%%%%%%%%%%%%%%%%%%

%%%%%%%%%%%%%%%%%%%%%
\author[10]{Jos{\' e}~Bernabeu,}
%%\email{andrea.bertoldi@institutoptique.fr}
\affiliation[10]{Department of Theoretical Physics, University of Valencia, and IFIC, Joint Centre Univ. Valencia-CSIC, E-46100 Burjassot, Valencia}
%%%%%%%%%%%%%%%%%%%%%%

%%%%%%%%%%%%%%%%%%%%%
\author[8,*]{Andrea~Bertoldi,}
%%\email{andrea.bertoldi@institutoptique.fr}
%\affiliation[8]{LP2N, Laboratoire Photonique, Num\'erique et Nanosciences,\\ Universit\'e Bordeaux-IOGS-CNRS:UMR 5298, F-33400 Talence, France.}
%%%%%%%%%%%%%%%%%%%%%%

%%%%%%%%%%%%%%%%%%%%%
\author[11]{Robert~Bingham,}
%%\email{andrea.bertoldi@institutoptique.fr}
\affiliation[11]{STFC, Rutherford Appleton Laboratory, Didcot OX110QX, UK}
%%%%%%%%%%%%%%%%%%%%%%

%%%%%%%%%%%%%%%%%%%%%
\author[12]{Diego~Blas,}
%%\email{andrea.bertoldi@institutoptique.fr}
\affiliation[12]{Department of Physics, King's College London, Strand, London WC2R 2LS, UK}
%%%%%%%%%%%%%%%%%%%%%%

%%%%%%%%%%%%%%%%%%%%%%%%%%%%%%
\author[13]{Vasiliki~Bolpasi,}
%\emailAdd{J.Coleman@liverpool.ac.uk}
\affiliation[13]{Institute of Electronic Structure and Laser, Foundation for Research and Technology-Hellas, Vassilika Vouton, Heraklion 70013, Greece}
%%%%%%%%%%%%%%%%%%%%%%%%%%

%%%%%%%%%%%%%%%%%%%%%
\author[14,*]{Kai~Bongs,}
%%\email{K.Bongs@bham.ac.uk}
%\affiliation[2]{School of Physics and Astronomy, University of Birmingham, UK}
\affiliation[14]{Department of Physics and Astronomy, University of Birmingham, Edgbaston, Birmingham B15 2TT, UK}
%%%%%%%%%%%%%%%%%%%%%%

%%%%%%%%%%%%%%%%%%%%%
\author[15]{Sougato~Bose,}
%%\email{andrea.bertoldi@institutoptique.fr}
\affiliation[15]{Department of Physics and Astronomy, University College London, London, UK}
%%%%%%%%%%%%%%%%%%%%%%

%%%%%%%%%%%%%%%%%%%%%%%%%%%%%%%%
\author[8,*]{Philippe~Bouyer,}
%%\email{philippe.bouyer@institutoptique.fr}
%\affiliation[a]{LP2N, Laboratoire Photonique, Num\'erique et Nanosciences,\\ Universit\'e Bordeaux-IOGS-CNRS:UMR 5298, F-33400 Talence, France.}
%%%%%%%%%%%%%%%%%%%%%%

%%%%%%%%%%%%%%%%%%%%%
\author[16]{Themis~Bowcock,}
%%\email{andrea.bertoldi@institutoptique.fr}
\affiliation[16]{Department of Physics, University of Liverpool, Liverpool L69 7ZE, UK}
%%%%%%%%%%%%%%%%%%%%%%

%%%%%%%%%%%%%%%%%%%%%
\author[17]{William~Bowden,}
%%\email{andrea.bertoldi@institutoptique.fr}
\affiliation[17]{National Physical Laboratory, Teddington TW11 0LW, UK}
%%%%%%%%%%%%%%%%%%%%%%

%%%%%%%%%%%%%%%%%%%%
\author[4,@]{Oliver~Buchmueller,}
\emailAdd{ o.buchmueller@imperial.ac.uk}
%\affiliation[4]{High Energy Physics Group, Blackett Laboratory, Imperial College, Prince Consort Road, London, SW7 2AZ, UK}
%%%%%%%%%%%%%%%%%%%%%%%%%%%%

%%%%%%%%%%%%%%%%%%%%%
\author[18]{Clare~Burrage,}
%%\email{andrea.bertoldi@institutoptique.fr}
\affiliation[18]{School of Physics and Astronomy, University of Nottingham, Nottingham NG7 2RD, UK}
%%%%%%%%%%%%%%%%%%%%%%

%%%%%%%%%%%%%%%%%%%%%
\author[19]{Xavier~Calmet,}
%%\email{andrea.bertoldi@institutoptique.fr}
\affiliation[19]{Department of Physics and Astronomy, University of Sussex, Brighton BN1 9QH, UK}
%%%%%%%%%%%%%%%%%%%%%%

%%%%%%%%%%%%%%%%%
\author[8,*]{Benjamin~Canuel,}
%%\email{benjamin.canuel@institutoptique.fr}
%\affiliation[a]{LP2N, Laboratoire Photonique, Num\'erique et Nanosciences,\\ Universit\'e Bordeaux-IOGS-CNRS:UMR 5298, F-33400 Talence, France.}
%%%%%%%%%%%%%%%%%%%%%%

%%%%%%%%%%%%%%%%%%%%%%%%%%%%%
\author[20,*]{Laurentiu-Ioan~Caramete,}
%\emailAdd{ lcaramete@spacescience.ro}
\affiliation[20]{Institute of Space Science, 409, Atomistilor Street Magurele, 077125 Ilfov, Romania}
%%%%%%%%%%%%%%%%%%%%%%%%%%%

%%%%%%%%%%%%%%%%%%%%%%%%%%%%%%%
\author[16]{Andrew~Carroll,}
%\emailAdd{J.Coleman@liverpool.ac.uk}
%\affiliation[20]{Department of Physics, University of Liverpool, Liverpool L69 7ZE, UK}
%%%%%%%%%%%%%%%%%%%%%%%%%%

%%%%%%%%%%%%%%%%%%%%%%%%%%%%%%%
\author[21,22]{Giancarlo~Cella,}
%\emailAdd{maria.luisa.chiofalo@unipi.it}
%\affiliation[21]{Dipartimento di Fisica``Enrico Fermi", Universi\`a di Pisa and INFN, Largo Bruno Pontecorvo 3, I-56126 Pisa, Italy}
\affiliation[21]{Dipartimento di Fisica``Enrico Fermi", Universit\`a di Pisa, Largo Bruno Pontecorvo 3, I-56126 Pisa, Italy}
\affiliation[22]{INFN, Sezione di Pisa, Largo Bruno Pontecorvo 3, I-56126 Pisa, Italy}
%%%%%%%%%%%%%%%%%%%%%%%%%%

%%%%%%%%%%%%%%%%%%%%%%%%%%%%%%%
\author[23]{Vassilis~Charmandaris,}
%\emailAdd{maria.luisa.chiofalo@unipi.it}
\affiliation[23]{University of Crete and Foundation for Research and Technology-Hellas, Heraklion, Greece}
%%%%%%%%%%%%%%%%%%%%%%%%%%
%%%%%%%%%%%%%%%%%%%%%%%%%%%%%%%
\author[24,25]{Swapan~Chattopadhyay,}
%\emailAdd{J.Coleman@liverpool.ac.uk}
\affiliation[24]{Department of Physics, Northern Illinois University, DeKalb, IL 60115, USA}
\affiliation[25]{Fermi National Accelerator Laboratory,
P.O. Box 500, Batavia, IL 60510-0500, USA}
%%%%%%%%%%%%%%%%%%%%%%%%%%

%%%%%%%%%%%%%%%%%%%%%%%%%%%%%%%
\author[26]{Xuzong~Chen,}
%\emailAdd{J.Coleman@liverpool.ac.uk}
\affiliation[26]{Department of Electronics, Peking University, Beijing 100871, China}
%%%%%%%%%%%%%%%%%%%%%%%%%%

%%%%%%%%%%%%%%%%%%%%%%%%%%%%%%%
\author[21,22]{Maria~Luisa~Chiofalo,}
%\emailAdd{maria.luisa.chiofalo@unipi.it}
%\affiliation[22]{Dipartimento di Fisica``Enrico Fermi", Universi\`a di Pisa and INFN, Largo Bruno Pontecorvo 3, I-56126 Pisa, Italy}
%%%%%%%%%%%%%%%%%%%%%%%%%%

%%%%%%%%%%%%%%%%%%%%%%%%%%%%%%%
\author[16,*]{Jonathon~Coleman,}
%\emailAdd{J.Coleman@liverpool.ac.uk}
%\affiliation[6]{University of Liverpool, Liverpool L69 7ZE, UK}
%%%%%%%%%%%%%%%%%%%%%%%%%%

%%%%%%%%%%%%%%%%%%%%%%%%%%%%%%%
\author[4]{Joseph~Cotter,}
%\emailAdd{J.Coleman@liverpool.ac.uk}
%\affiliation[4]{Northern Illinois University, USA}
%%%%%%%%%%%%%%%%%%%%%%%%%%

%%%%%%%%%%%%%%%%%%%%%%%%%%%%%%%
\author[27]{Yanou~Cui,}
%\emailAdd{J.Coleman@liverpool.ac.uk}
\affiliation[27]{Department of Physics and Astronomy, University of California, Riverside, CA 92521-0413, USA}
%%%%%%%%%%%%%%%%%%%%%%%%%%

%%%%%%%%%%%%%%%%%%%%%%%%%%%%%%%
% ...
%%%%%%%%%%%%%%%%%%%%%%

\maketitle
%\flushbottom
%%%%%%%%%%%%%%%%%%%%%%%%%%%%%%%%%%%%%%%%%%%%%%%%%%%
%\section{Introduction} 
%%%%%%%%%%%%%%%%%%%%%%%%%%%%%%%%%%%%%%%%%%%%%%%%%%%

\section*{Comment for arXiv} 
This paper is based on a submission (v1) in response to the Call for White Papers for the Voyage 2050 long-term plan in the ESA Science Programme. ESA limited the number of White Paper authors to 30. However, in this version (v2) we have welcomed as supporting authors participants in the Workshop on Atomic Experiments for Dark Matter and Gravity Exploration held at CERN: ({\tt https://indico.cern.ch/event/830432/}), as well as other interested scientists, and have incorporated additional material. 

\newpage

%%%%%%%%%%%%%%%%%%%%%%%%%%%%%%%%%%%%%%%%%%%%%%%%%%%
%\section{Introduction} 
%%%%%%%%%%%%%%%%%%%%%%%%%%%%%%%%%%%%%%%%%%%%%%%%%%%
\section{Preface} 

This article originates from the {\it Workshop on Atomic Experiments for Dark Matter and Gravity Exploration}~\cite{workshop}, which took place on July 22 and 23, 2019, hosted by CERN, Geneva, Switzerland.

This workshop reviewed the landscape of cold atom technologies being developed to explore fundamental physics, astrophysics and cosmology - notably ultra-light dark matter and gravitational effects, particularly gravitational waves in the mid-frequency band between the maximal sensitivities of existing and planned terrestrial and space experiments, and searches for new fundamental interactions - which offer several opportunities for ground-breaking discoveries.  

The goal of the workshop was to bring representatives of the cold atom community together with colleagues from the particle physics and gravitational communities, with the aim of preparing for ESA the White Paper that is the basis for this article. It outlines in Sections~\ref{SC} and \ref{physics} the science case for a future space-based cold atom detector mission discussed in Sections~\ref{MC}, based on technologies described in Section~\ref{TR}, and is summarized in Section~\ref{summary}.

%\section{Introduction}
%{\it add intro ref. to sections }
\section{Science Case}
\label{SC}

Two of the most important issues in fundamental physics, astrophysics and cosmology
are the nature of dark matter (DM) and the exploration of the gravitational wave (GW) spectrum.

Multiple observations from the dynamics of galaxies and clusters to the spectrum of the
cosmological microwave background (CMB) radiation measured by ESA's Planck satellite
and other~\cite{Aghanim:2018eyx} experiments indicate that there is far more DM than
conventional matter in the Universe, but its physical composition remains a complete mystery.
The two most popular classes of DM scenario invoke either 
coherent waves of ultra-light bosonic fields,
or weakly-interacting massive particles (WIMPs). In the absence so far of any positive indications
for WIMPs from accelerator and other laboratory experiments, there is increasing interest in
ultra-light bosonic candidates, many of which appear in theories that address other problems in
fundamental physics. {\it Such bosons are among the priority targets for AEDGE.}

The discovery of GWs by the LIGO~\cite{TheLIGOScientific:2014jea} and Virgo~\cite{TheVirgo:2014hva} laser interferometer
experiments has opened a new window on the Universe, through which waves over a wide range of frequencies can provide new information about
high-energy astrophysics and cosmology. Just as astronomical observations at different wavelengths
provide complementary information about electromagnetic sources, measurements of GWs in different
frequency bands are complementary and synergistic. In addition to the ongoing LIGO and Virgo
experiments at relatively high frequencies $\gtrsim 10$~Hz, which will soon be joined by KAGRA~\cite{Somiya:2011np} and
INDIGO~\cite{Unnikrishnan:2013qwa}, with the Einstein Telescope (ET)~\cite{Punturo:2010zz,Sathyaprakash:2012jk} and Cosmic Explorer (CE)~\cite{Reitze:2019iox} experiments being planned for similar frequency ranges, ESA has approved for launch before
the period being considered for Voyage 2050 missions the LISA mission, which will be most sensitive
at frequencies $\lesssim 10^{-1}$~Hz, and the Taiji~\cite{Guo:2018npi} and TianQin~\cite{Luo_2016} missions proposed in China will have similar sensitivity to LISA. AEDGE is optimized for the mid-frequency range between
LISA/Taiji/TianQin and LIGO/Virgo/KAGRA/INDIGO/ET/CE~\footnote{The ALIA proposal in Europe~\cite{Bender:2013nsa} and the DECIGO proposal in Japan~\cite{Kawamura:2011zz} have been aimed at a similar frequency range, and the scientific interest of this frequency range has recently been stressed in~\cite{Mandel:2017pzd,Baker:2019pnp} and~\cite{Kuns:2019upi}.}. This range is ideal for probing the formation of the super-massive black holes known to be present in many galaxies.  Also, AEDGE's observations of astrophysical sources will complement those by  other GW experiments at lower and higher frequencies, completing sets of measurements from inspiral to merger and ringdown, yielding important
synergies as we illustrate below. {\it GWs are the other priority targets for AEDGE.}

In addition to these primary scientific objectives, several other potential objectives
for cold atom experiments in space are under study. These may include searches for
astrophysical neutrinos, constraining possible variations in fundamental constants,
probing dark energy, and probing basic physical principles such as Lorentz invariance
and quantum mechanics. Cold quantum gases provide powerful technologies that are already mature for the AEDGE goals, while also developing rapidly~\cite{Pezze:2018wyn}. The developments of these technologies can be expected to offer AEDGE more possibilities on the Voyage 2050 time scale. {\it AEDGE is a uniquely interdiscplinary and versatile mission.}

An atom interferometer such as AEDGE is sensitive to fluctuations in the relative phase between cold atom clouds separated by a distance $L$:
\begin{equation}
\Delta \phi \; = \; \omega_A \times (2 L) \, ,
\label{basic}
\end{equation}
where $\omega_A$ is the frequency of the atomic transition being studied.
DM interactions with the cold atoms could induce variations $\delta \omega_A$ in this frequency, and the passage of a GW inducing a strain $h$ would induce a phase shift via a change $\delta L = h L$ in the distance of separation. The AEDGE capabilities for DM detection are summarized in Section~\ref{ULDM}, where we show how AEDGE can explore the
parameters of ultra-light DM models orders of magnitude beyond current bounds. The AEDGE capabilities for
GW measurements are discussed in Section~\ref{GW}, where we
stress its unique capabilities for detecting GWs from the mergers of intermediate-mass black holes, as well as from first-order phase transitions in the early universe and cosmic strings. Finally, AEDGE prospects for other fundamental physics topics
are outlined in Section~\ref{FP}. One specific measurement concept is described in
Section~\ref{MC}, but other concepts can be considered, as reviewed in Section~\ref{TR}.
The cold atom projects mentioned there may be considered as ``pathfinders" for the AEDGE mission, providing a roadmap towards its realization that is outlined in Section~\ref{summary}. These experiments include many terrestrial
cold atom experiments now being prepared or proposed, space experiments such as cold atom
experiments on the ISS, LISA Pathfinder and LISA itself. With this roadmap in mind, the AEDGE concept is being proposed
by experts in the cold atom community, as well as GW experts and fundamental particle physicists.

\section{AEDGE Capabilities for its Scientific Priorities}
\label{physics}
In this section we develop the science case of AEDGE, providing important examples of its capabilities for its primary scientific objectives, namely the DM search and GW detection, and mentioning also other potential science topics. The basis of the sensitivity projections shown here is defined in Section~\ref{MC}.

\subsection{Dark Matter}
\label{ULDM}

Multiple observations point to the existence of dark matter (DM), an elusive form of matter that comprises around 84\% of the matter energy density in the Universe~\cite{Aghanim:2018eyx}. So far, all of the evidence for DM arises through its gravitational interaction, which provides little insight into the DM mass, but it is anticipated that DM also interacts with normal matter through interactions other than gravity.

The direct search for DM, which aims to detect the non-gravitational interaction of DM in the vicinity of the Earth, is one of the most compelling challenges in particle physics. The direct search for DM in the form of an (electro-)weakly-interacting massive particle (WIMP) with a mass in the GeV to multi-TeV window is mature, and experiments now probe interaction cross-sections far below the electroweak scale. As yet, no positive detections have been reported (see e.g., the constraints from XENON1T~\cite{Aprile:2018dbl}),
and the same is true of collider searches for WIMPs and indirect searches among cosmic rays and $\gamma$ rays for the products
of annihilations of astrophysical WIMPs. Although the experimental search for electroweak-scale DM has been the most prominent, theoretical extensions of the Standard Model (SM) of particle physics provide many other elementary particle candidates for DM over a much wider mass scale: ranging from $10^{-22}$~eV to the Planck scale $\sim 10^{18}$~GeV~\cite{Battaglieri:2017aum}. 

Ultra-light DM (with a sub-eV mass) is particularly interesting, as there are many well-motivated candidates. These include the QCD axion and axion-like-particles (ALPs); (dark) vector bosons; and light scalar particles such as moduli, dilatons or the relaxion. Ultra-light bosons are also good DM candidates: there are well-understood mechanisms to produce the observed abundance (e.g., the misalignment mechanism~\cite{Preskill:1982cy, Abbott:1982af, Dine:1982ah}), and the DM is naturally cold, so it is consistent with the established structure formation paradigm.
%~~\\

\mysubsubsection{~~~~~~~~~~~~~~~~~~~~~~~~~~~~~~~~~~~~~~~~~~~~~~~~~~~~~~~~~~~~~Scalar dark matter}
%~~\\

%\vspace{-0.5cm}
\begin{wrapfigure}[32]{lt}{0.5\textwidth}
%\begin{figure}[t!]
%\centering
~~\\
\vspace{-2.2cm}
\includegraphics[width=0.475\textwidth]{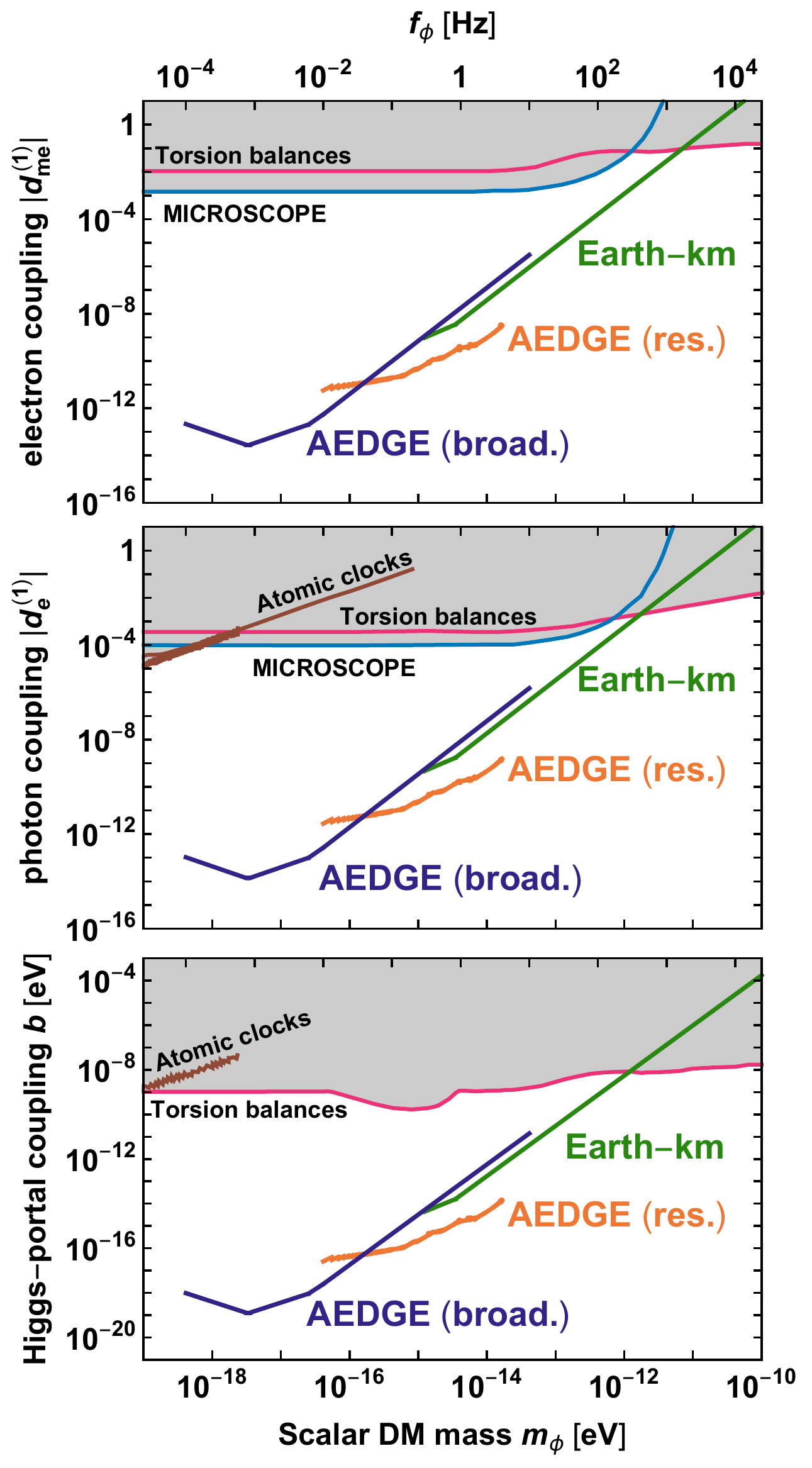}
\vspace{1.6cm}
\caption{\it The sensitivities of AEDGE in broadband (purple lines) and resonant mode (orange lines) to linear scalar DM interactions with electrons (top), photons (middle) and via the Higgs portal (bottom), compared to those of a km-scale terrestrial experiment (green lines). The grey regions show parameter spaces that have been excluded by the MICROSCOPE experiment (blue lines)~\cite{Berge:2017ovy,Hees:2018fpg}, searches for violations of the equivalence principle with torsion balances (red lines)~\cite{Schlamminger:2007ht,Wagner:2012ui}, or by atomic clocks (brown lines)~\cite{VanTilburg:2015oza, Hees:2016gop}.}
\label{DMplot}
\vspace{-110pt}
\end{wrapfigure}
%\vspace{-0.2cm}
Atom interferometers are able to measure a distinctive signature of scalar DM~\cite{Geraci:2016fva, Arvanitaki:2016fyj}. Scalar DM may cause fundamental parameters such as the electron mass and electromagnetic fine-structure constant to oscillate in time, with a frequency set by the mass of the scalar DM and an amplitude determined by the DM mass and local DM density~\cite{Arvanitaki:2014faa,Stadnik:2015kia}. This in turn leads to a temporal variation of atomic transition frequencies, since the transition frequencies depend on the electron mass and fine-structure constant. A non-trivial signal phase occurs in a differential atom interferometer when the period of the DM wave matches the total duration of the interferometric sequence~\cite{Arvanitaki:2016fyj}.

We consider first scenarios where scalar DM couples {\it linearly} to the Standard Model fields~\cite{Damour:2010rm,Damour:2010rp} through an interaction of the form
\begin{equation}
\begin{split}
\mathcal{L}^{\rm{lin}}_{\rm{int}}&\supset - \phi \sqrt{4 \pi G_{\rm{N}}}  \left[ d^{(1)}_{me} m_e \bar{e}e - \frac{d^{(1)}_e}{4} F_{\mu \nu} F^{\mu \nu} \right] \\
& \quad +b\, \phi |H|^2    
\label{linear}
\end{split}
\end{equation}
Fig.~\ref{DMplot} shows the projected sensitivity of AEDGE for three scenarios:
light scalar DM with a coupling $d^{(1)}_{me}$ to electrons (top), a coupling $d^{(1)}_e$ to photons (middle), and a Higgs-portal coupling $b$ (bottom). The coloured lines show the couplings that can be detected at signal-to-noise (SNR) equal to one after an integration time of $10^8$~s.  We show predictions for AEDGE operating in broadband (purple lines) and resonant mode (orange lines) with the sensitivity parameters given in Table~\ref{tab:parameters} below.

The sensitivity of AEDGE in broadband mode extends from $\sim 10^2$ down to $\sim 10^{-4}$~Hz, which is the approximate frequency where gravity gradients become more important than shot noise~\cite{Arvanitaki:2016fyj}. Also shown for comparison are
the sensitivities of a km-scale ground-based interferometer scenario.\footnote{This projection assumes that the gravity gradient noise (GGN) can be mitigated, as discussed later.}
The grey regions show parameter space that has already been excluded by the indicated experiments. 
We see that AEDGE will probe extensive new regions of parameter space for the electron coupling,
extending down to $\sim 10^{-14}$ for a scalar mass $\sim 10^{-17}$~eV, and similarly for
a photon coupling, while the sensitivity to a Higgs-portal coupling would extend down to $10^{-19}$~eV for this mass. We see also that the sensitivities of AEDGE would extend to
significantly lower masses and couplings than a possible km-scale terrestrial experiment, used here as a benchmark. 
%The sensitivity projections shown here assume that the interferometer runs in broadband mode so that it is sensitive to a wide range of scalar DM masses. However, it is also possible to operate the interferometer in resonant mode~\cite{Graham:2016plp}. I
Fig.~\ref{DMplot} also shows that when operated in resonant mode AEDGE will have extended sensitivity between $10^{-16}$~eV and $10^{-14}$~eV: see Ref.~\cite{Arvanitaki:2016fyj} for further details.

%\begin{figure}[h!]
%\hspace{-5mm}
%\centering
%\includegraphics[width=1\textwidth]{DM_AEDGE.pdf}
%\caption{\it The sensitivity of AEDGE to scalar DM interactions with electrons (left), photons (middle) and through the Higgs portal (right), compared to that of a possible 1-km terrestrial experiment. The grey regions show parameter space that has already been excluded through searches for violations of the equivalence principle~\cite{Wagner:2012ui}, atomic spectroscopy~\cite{Hees:2016gop} or by the MICROSCOPE experiment~\cite{Berge:2017ovy}. {\color{red} Needs updating of present constraints and addition of resonant sensitivity by Chris.}}
%\label{DMplot}
%\end{figure}

\begin{figure}[t]
\centering
\includegraphics[width=0.475\textwidth]{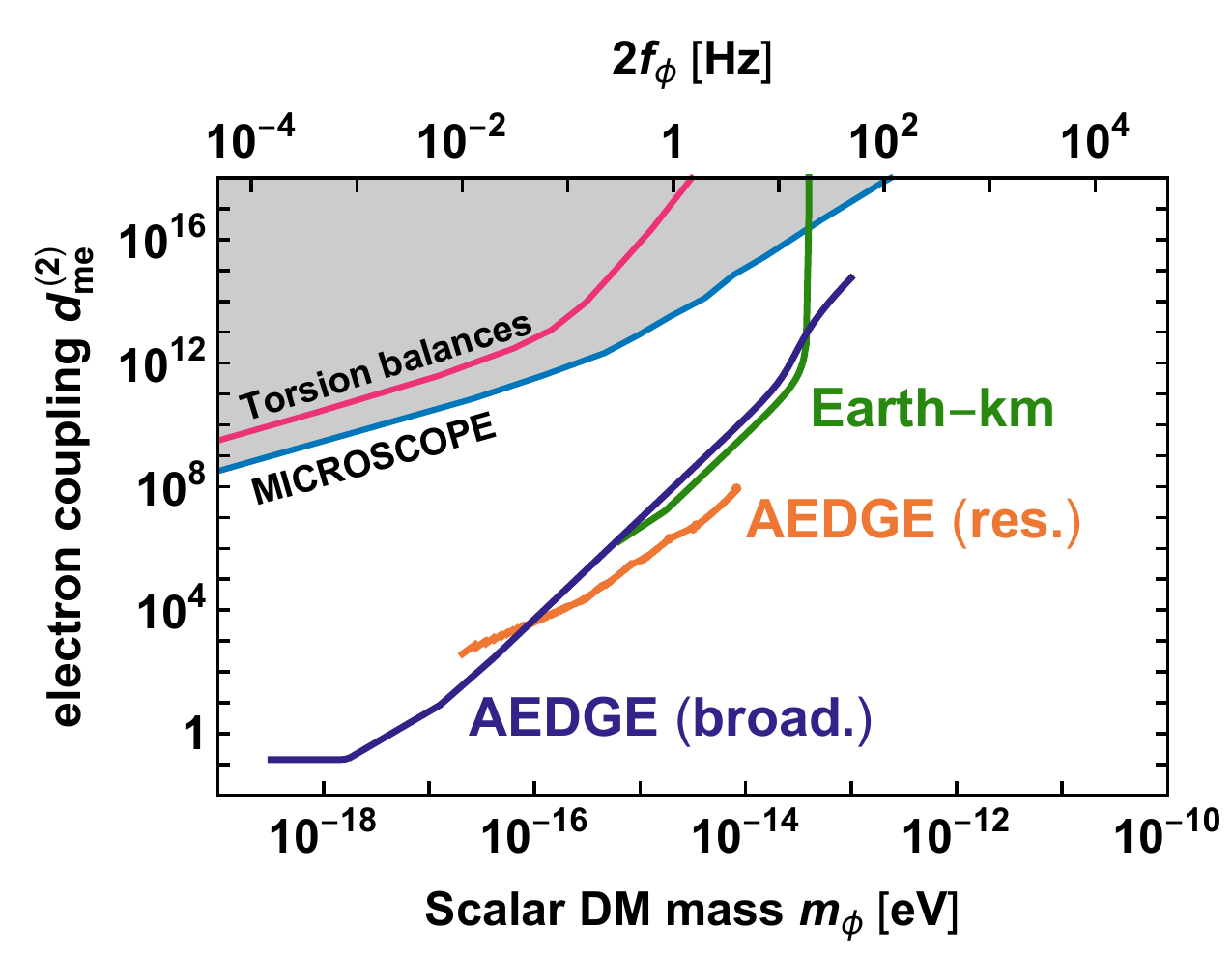} \hspace{4mm}
\includegraphics[width=0.475\textwidth]{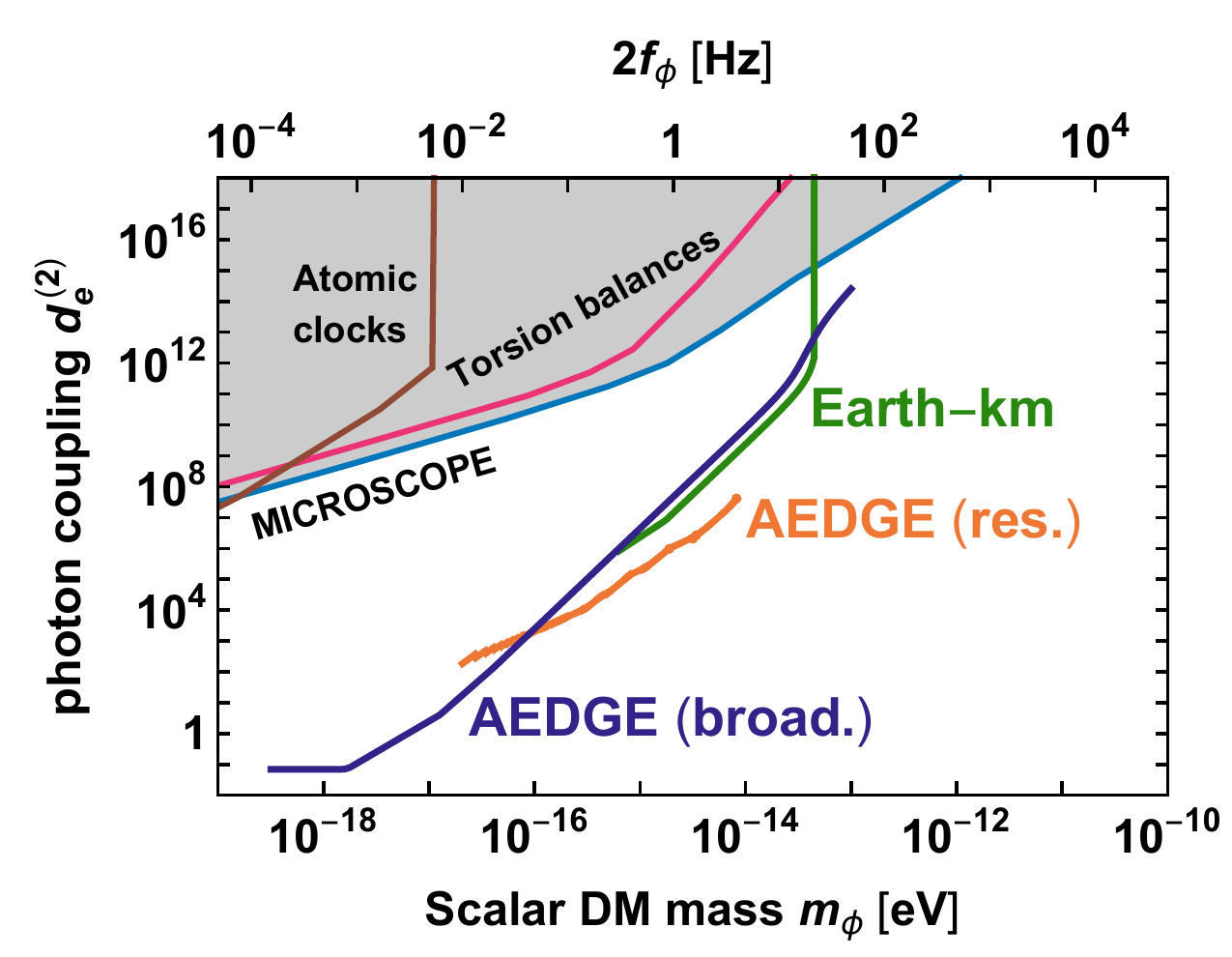}
\caption{\it The sensitivities of AEDGE in broadband (purple lines) and resonant mode (orange lines) to quadratic scalar DM interactions with electrons (left) and photons (right), compared to those of a km-scale terrestrial experiment (green lines). The grey regions show parameter spaces that have been excluded by the MICROSCOPE experiment (blue lines)~\cite{Berge:2017ovy,Hees:2018fpg}, searches for violations of the equivalence principle with torsion balances (red lines)~\cite{Schlamminger:2007ht,Wagner:2012ui}, or by atomic clocks (brown lines)~\cite{VanTilburg:2015oza,Hees:2016gop}.}
\label{DMplotquad}
\end{figure}

Fig.~\ref{DMplotquad} illustrates AEDGE capabilities in a scenario where scalar DM couples {\it quadratically} to Standard Model fields~\cite{Stadnik:2014tta}: 
\begin{equation}
\mathcal{L}^{\rm quad}_{\rm{int}} \supset - \;\phi^2  \cdot 4 \pi G_{\rm{N}} \cdot  \left[ d^{(2)}_{ me}\,
m_e  \bar{e} e \; + \; \frac{d^{(2)}_{\rm e}}{4} F_{\mu \nu} F^{\mu \nu} \right] \, .
\label{quadratic}
\end{equation}
Limits and sensitivities to the quadratic coupling $d^{(2)}_{me}$ of scalar DM to electrons are shown in the left panel of Fig.~\ref{DMplotquad}, and those for a quadratic coupling $d^{(2)}_e$ to photons in the right panel.\footnote{It has been pointed out in~\cite{Stadnik:2014tta} that, in  addition to the constraints displayed in Fig.~\ref{DMplotquad}, there are are potential constraints on quadratically-coupled DM from Big Bang Nucleosynthesis, which merit detailed evaluation.}
As in Fig.~\ref{DMplot}, the coloured lines show the couplings that can be detected at SNR equal to one for AEDGE operating in broadband (purple lines) and resonant mode (orange lines). 
We see that AEDGE will also probe extensive new regions of parameter space for the electron and photon quadratic couplings,
extending the sensitivity to values of $d^{(2)}_{me}$ and $d^{(2)}_e$ by up to eight orders of magnitude. The quadratic couplings give rise to a richer phenomenology than that offered by linear couplings. For example, a screening mechanism occurs for positive couplings, which reduces the sensitivity of terrestrial experiments~\cite{Hees:2018fpg}. This is illustrated in Fig.~\ref{DMplotquad} by the steep rises in the atomic clock constraints and the sensitivity of a km-scale ground-based interferometer. By comparison, space-based experiments are less affected by the screening mechanism and AEDGE therefore maintains sensitivity at larger masses.

As outlined in~\cite{Geraci:2016fva}, AEDGE could also be sensitive to additional ranges of scalar DM masses via direct accelerations of the atoms produced by interactions with dark matter fields, and also through the indirect effects of the inertial and gravitational implications of the variations of the atomic masses and the mass of the Earth. It is estimated that several orders of magnitude of additional unexplored phase space for DM couplings in the mass range of $\sim 10^{-2}$~eV to $\sim 10^{-16}$~eV could be probed via these new effects.
~~\\

%\newpage
\mysubsubsection{Axion-like particles and vector dark matter}
~\\

In addition to scalar dark matter, atom interferometers can search for other ultra-light DM candidates.

$\bullet$ Axion-like DM causes the precession of nuclear spins around the axion field. Using atomic isotopes with different nuclear spins, atom interferometers are sensitive to the axion-nucleon coupling for axion-like DM lighter than $10^{-14}$~eV~~\cite{Graham:2017ivz}.

$\bullet$ Two interferometers running simultaneously with two different atomic species act as an accelerometer. This set-up is sensitive to, for instance, a dark vector boson with a mass below $10^{-15}$~eV coupled to the difference between baryon number (B) and lepton number (L)~\cite{Graham:2015ifn}.
~~\\

\mysubsubsection{Identifying a DM signal}
~\\

Confirming that the origin of a positive detection is due to a DM signal may be challenging. However, there are a number of characteristic features of the DM signal that should allow it to be distinguished from other sources. For example, compared to GW signals from binary systems, where the frequency changes as the binary system evolves, the frequency of the DM signal is set by the mass of the scalar DM and will therefore remain constant. The DM speed distribution may also have distinctive features (see e.g.,~\cite{OHare:2018trr}) 
and there is a characteristic modulation over the course of a year, caused by the rotation of the Earth about the Sun~\cite{Roberts:2018agv}. If these distinctive features can be measured, they would point to a DM origin for the signal.  

\subsection{Gravitational Waves}
\label{GW}

The first direct evidence for gravitational waves (GWs) came from the LIGO/Virgo discoveries of emissions
from the mergers of black holes (BHs) and of neutron stars~\cite{LIGOScientific:2018mvr}. These discoveries open new vistas in the
exploration of fundamental physics, astrophysics and cosmology. Additional GW experiments are now being prepared
and proposed, including upgrades of LIGO~\cite{TheLIGOScientific:2014jea} and Virgo~\cite{TheVirgo:2014hva}, KAGRA~\cite{Somiya:2011np}, INDIGO~\cite{Unnikrishnan:2013qwa}, the Einstein Telescope (ET)~\cite{Punturo:2010zz,Sathyaprakash:2012jk}
and Cosmic Explorer (CE)~\cite{Reitze:2019iox}, which will provide
greater sensitivities in a similar frequency range to the current LIGO and Virgo experiments, and LISA~\cite{Audley:2017drz}, which
will provide sensitivity in a lower frequency band on a longer time-scale. 
In addition, pulsar timing arrays provide sensitivity to GWs in a significantly lower
frequency band~\cite{vanHaasteren:2011ni}.

As we discuss in more detail below, there are several terrestrial cold atom experiments that are currently 
being prepared, such as MIGA~\cite{Canuel:2017rrp}, ZAIGA~\cite{Zhan:2019quq} and MAGIS~\cite{Graham:2017pmn}, 
or being proposed, such as ELGAR~\cite{Bouyer} and AION~\cite{AION}. These experiments will
provide measurements complementary to LISA and LIGO/Virgo/KAGRA/INDIGO/ET/CE
via their sensitivities in the mid-frequency range between 1 and $10^{-2}$~Hz.

AEDGE will subsequently provide a significantly extended reach for 
GWs in this frequency range, as we illustrate in the following
with examples of astrophysical and cosmological sources of GWs,
which open up exciting new scientific opportunities.
~~\\

\mysubsubsection{Astrophysical Sources}
~~\\

The BHs whose mergers were discovered by LIGO and Virgo have masses up to several tens of solar masses.
On the other hand, supermassive black holes (SMBHs) with masses $> 10^6$ solar masses have been established as key ingredients in most if not all galaxies~\cite{Magorrian:1997hw}, and play major roles in cosmological structure 
formation and determining the shape, appearance and evolution of galaxies~\cite{Kauffmann:1999ce}. A first radio image of the SMBH in M87 has been released by the Event Horizon telescope (EHT)~\cite{Akiyama:2019cqa}, 
and observations of the Sgr A* SMBH at the centre of our galaxy are expected shortly.
The LISA frequency range is ideal for observations of mergers of SMBHs.

However, the formation and early evolution of SMBHs~\cite{Rees:1984si} and their possible connections to their  stellar mass cousins are still among the major unsolved puzzles in galaxy formation.  
It is expected that intermediate-mass black holes (IMBHs) with masses in the range 100 to $10^5$ solar masses
must also exist, and there is some observational evidence for them~\cite{Mezcua:2017npy}. They may well have played key roles in
the assembly of SMBHs. Detecting and characterising the mergers of IMBHs with  several hundred to a hundred thousand solar masses will provide evidence whether (and how) some 
of  the most massive ``stellar" black holes eventually grow into SMBHs~\cite{KSH}
or whether SMBHs grow from massive seed black holes formed by direct collapse from gas clouds in a  subset of low-mass galaxies~\cite{Volonteri:2002vz,Volonteri:2007ax}. 
\begin{figure}
\centering
%\vspace{-0.5cm}
\includegraphics[width=0.85\textwidth]{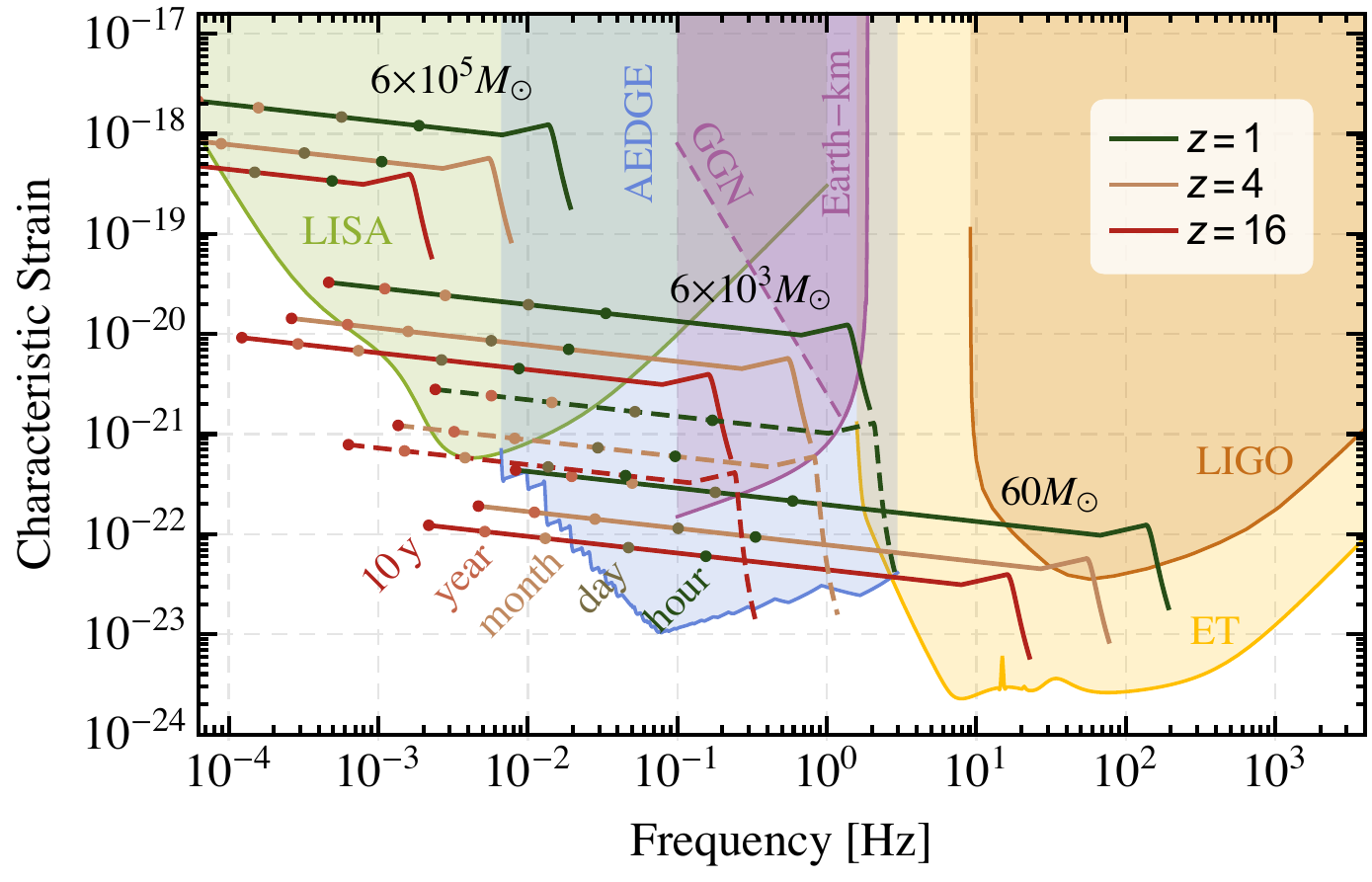}
\caption{\it Comparison of the strain measurements possible with AEDGE and other experiments, showing their sensitivities to BH mergers of differing total masses at various redshifts $z$, indicating also the time remaining before the merger. The solid lines correspond to equal mass binaries and the dashed ones to binaries with very different masses, namely $3000M_\odot$ and $30M_\odot$. Also shown is the possible gravitational gradient noise (GGN) level for a km-scale terrestrial detector, which would need to be mitigated for its potential to be realized. This figure illustrates the potential for synergies between AEDGE and detectors observing other stages of BH infall and merger histories.}
\label{staroplot}
\end{figure}

The AEDGE frequency range between $\sim 10^{-2}$ and a few Hz, where the LISA and the LIGO/Virgo/ KAGRA/INDIGO/ET/CE experiments are relatively insensitive, is ideal for observations of mergers involving IMBHs, as seen in Fig.~\ref{staroplot}. This figure shows that AEDGE {(assumed here to be operated in resonant mode)} would be able to observe the mergers of $6 \times 10^3$ solar-mass black holes out to very large redshifts $z$, as well as early inspiral stages of mergers of lower-mass BHs {of $60 M_\odot$, extending significantly the capabilities of terrestrial detectors to earlier inspiral stages. The dashed lines illustrate the observability of binaries with very different masses, namely $3000M_\odot$ and $30M_\odot$, which could be measured during inspiral, merger and ringdown phases out to large redshifts}~\footnote{This figure also indicates a typical gravitational gradient noise (GGN) level for a km-scale ground-based detector. In order for such a detector to reach its potential, this GGN would need to be significantly mitigated. Thanks to precise characterization of GGN correlation properties ~\cite{Junca:2019xvb}, it is possible to reduce GGN using detectors geometries based on arrays of Atom interferometers~\cite{Chaibi:2016dze}. A similar GGN level in a km-scale ground-based detector is relevant for the other GW topics discussed below.}.

\begin{figure}[h]
\centering
\includegraphics[width=0.45\textwidth]{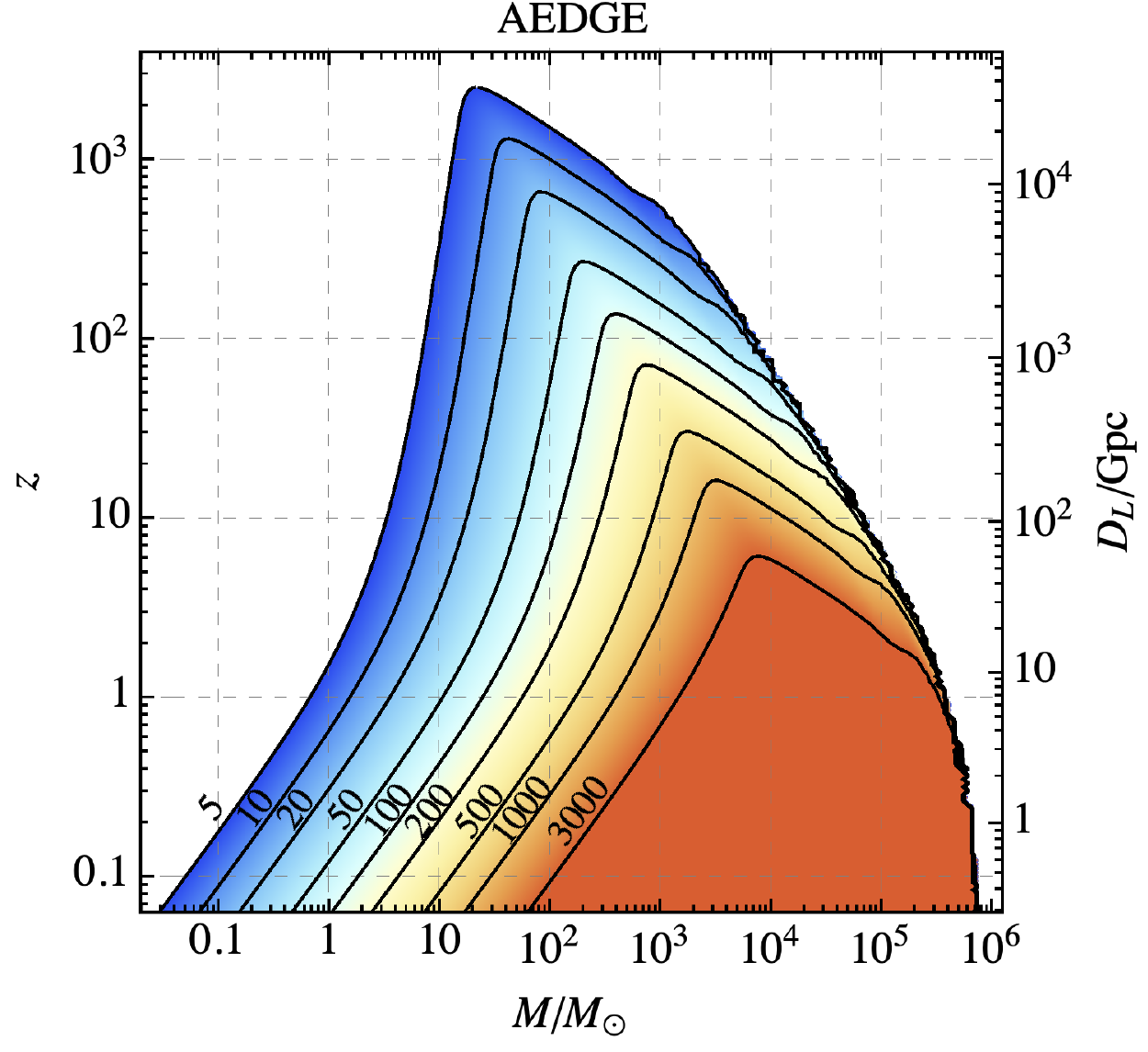} \hspace{4mm}
\includegraphics[width=0.45\textwidth]{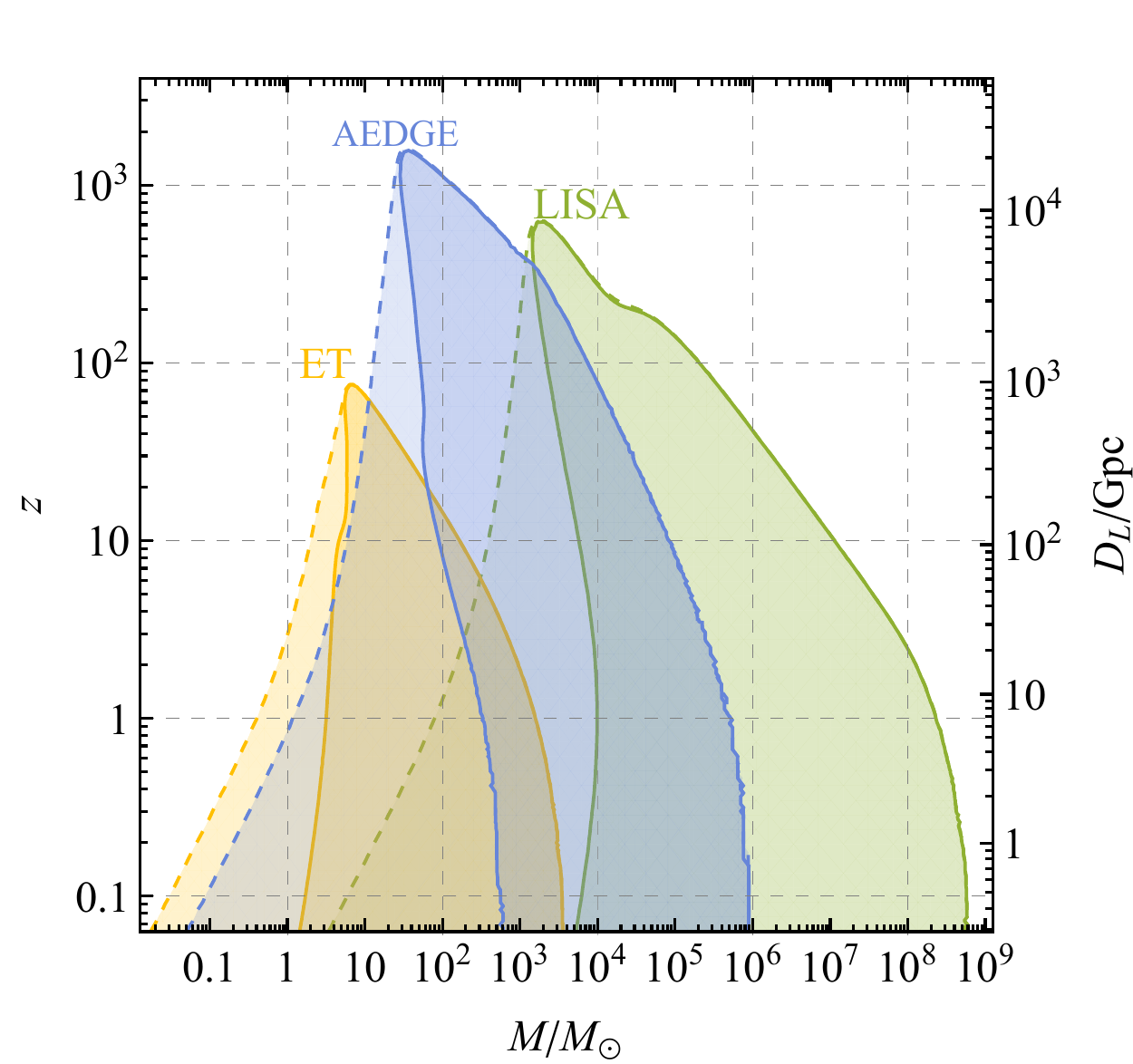}
\vspace{-2mm}
\caption{\it Left panel: The sensitivity of AEDGE to the mergers of IMBHs with the contours showing the signal-to-noise ratio (SNR). Right panel: Comparison of the sensitivities of AEDGE, ET and LISA with threshold ${\rm SNR}=8$. {In the lighter regions between the dashed and solid lines the corresponding detector observes only the inspiral phase.}}
\label{staroplot2}
\end{figure}

The left panel of Fig.~\ref{staroplot2} shows the sensitivity of AEDGE {operating in resonant mode} for detecting GWs from the mergers of IMBHs of varying masses at various signal-to-noise (SNR)
levels $\ge 5$. It could detect mergers of $\sim 10^4$ solar-mass BHs with SNR $\gtrsim 1000$ out to $z \sim 10$, where several dozen such events are expected per year according to~\cite{Erickcek:2006xc}, and mergers of $\sim 10^3$ solar-mass BHs with SNR $\gtrsim 100$ out to $z \gtrsim 100$. Such sensitivity should be sufficient to observe several hundred astrophysical BH mergers according to~\cite{Erickcek:2006xc}. This paper suggests that such events would be expected in the smaller part of this redshift range, so the observation of additional mergers at large redshifts could be a distinctive signature of primordial BHs.

{Another astrophysical topic where AEDGE can make a unique contribution is whether there is a gap in the spectrum of BH masses around $200 M_{\odot}$. We recall that electron-positron pair-instability is calculated to blow apart low-metallicity stars with masses around this value, leaving no BH remnant (see, for example,~\cite{Heger:2002by}). The AEDGE frequency range is ideal for measuring the inspirals of BHs with masses $\sim 200 M_{\odot}$ prior to their mergers. If they are observed, such BHs might be primordial, or come from higher-metallicity progenitors that are not of Population III, or perhaps have been formed by prior mergers.}

In addition to the stand-alone capabilities of AEDGE illustrated in Figs.~\ref{staroplot} and \ref{staroplot2}, there are significant synergies between AEDGE
measurements and observations in other frequency ranges, like those proposed in~\cite{Sesana2016} for the synergistic operation of LISA and LIGO: 

$\bullet$ 
{The measurement of early inspiral stages of BH-BH mergers of the type discovered by LIGO and Virgo is guaranteed science for AEDGE.} As seen in Fig.~\ref{staroplot}, AEDGE would observe out to high redshifts early inspiral stages of such mergers, which
could subsequently be measured weeks or months later by LIGO/Virgo/KAGRA/INDIGO/ET/CE. The inspiral phases of these sources could be measured for a month or more by AEDGE, enabling the times of subsequent mergers to be predicted accurately. The motion of the detectors around the Sun as well as in Earth orbit would make possible the angular localization with high precision of the coming merger~\cite{Graham:2017lmg}, providing `early warning' of possible upcoming multimessenger events. {The right panel of Fig.~\ref{staroplot2} compares the sensitivities of AEDGE at the SNR = 8 level (blue shading) with that of ET (yellow shading). The overlaps between the sensitivities show the possibilities for synergistic observations, with AEDGE measuring GWs emitted during the inspiral phase (lighter shading), and ET subsequently observing infall, the merger itself and the following ringdown phase (darker shading).}

\begin{figure}[h]
\centering
\includegraphics[width=\textwidth]{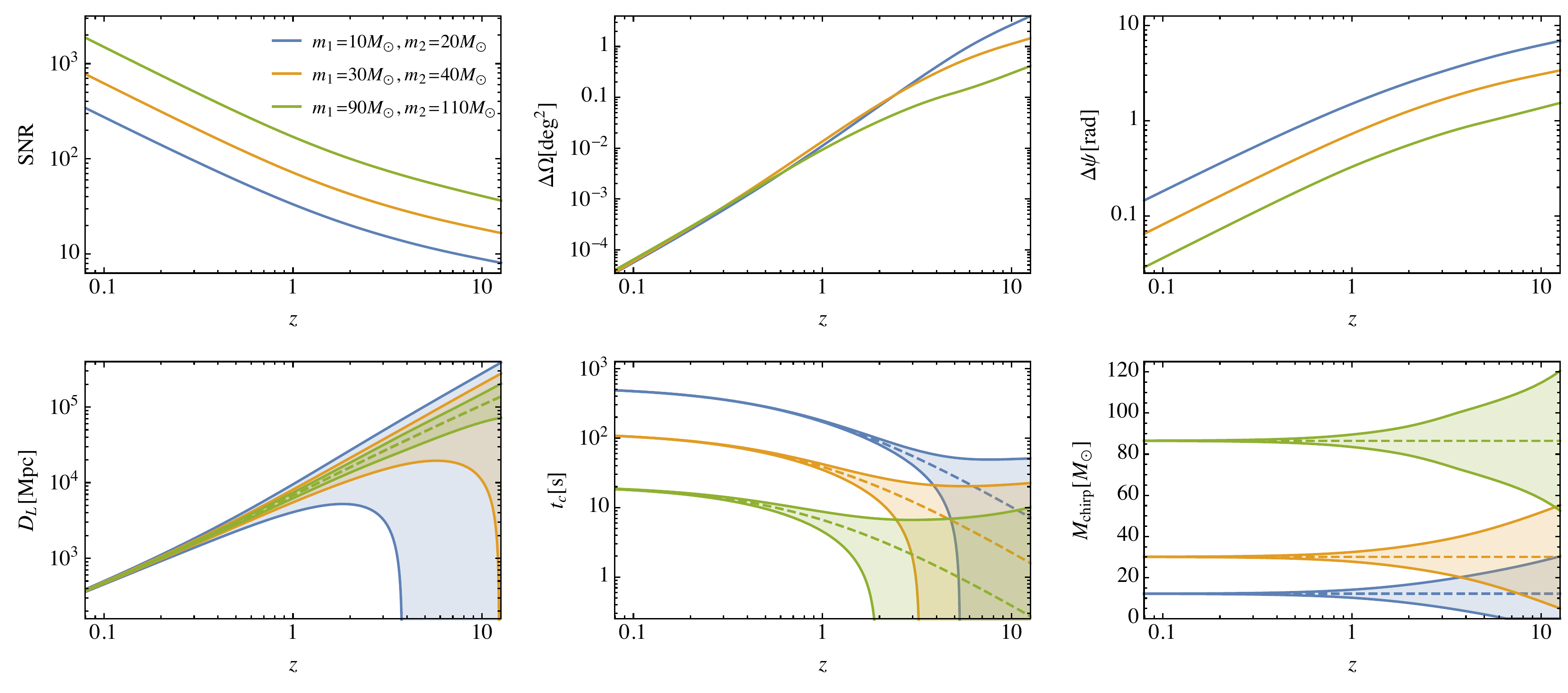}
\vspace{-6mm}
\caption{\it The SNR (upper left panel), the sky localization uncertainty $\Delta \Omega$ (upper middle panel), the polarization uncertainty $\Delta \psi$ (upper right panel), and the uncertainties in the luminosity distance $D_L$ (lower left panel), the time remaining before merger $t_c$ (lower middle panel) and the chirp mass $M_{\rm chirp}$ (lower right panel), calculated for three merging binaries of different BH mass combinations as functions of their redshifts.}
\label{fig:localization}
\end{figure}

{Fig.~\ref{fig:localization} shows some examples of these possible synergies for AEDGE measurements of the inspiral phases of binaries that merge in the LIGO/Virgo sensitivity window. The upper left plot shows the SNR as a function of redshift, and the other plots show how precisely various observables can be measured by observing for 180 days before the frequency of the signal becomes higher than 3\,Hz, corresponding to the upper limit of the AEDGE sensitivity window. As examples, we see in the upper middle panel that for events typical of those observed by LIGO/Virgo at $z\simeq 0.1$ the AEDGE sky localization uncertainty is less than $10^{-4}\,{\rm deg}^2$, while the upper right panel shows that the GW polarization could be measured accurately. The lower middle panel shows that the times of the mergers could be predicted with uncertainties measured in minutes, permitting advance preparation of comprehensive multimessenger follow-up campaigns. We also see in the lower panels that for binaries at high redshifts $z\gtrsim 1$ the uncertainties in the luminosity distance, the time before merger and the chirp mass become significant, though in these cases the measurements could be improved by starting to observe the binary more than 180 days before it exits the sensitivity window.}

$\bullet$ Conversely, as also seen in Fig.~\ref{staroplot} and the right panel of Fig.~\ref{staroplot2}, operating AEDGE within a few years of LISA would provide valuable synergies, as LISA
observations of inspirals (lighter green shading) could be used to make accurate predictions for subsequent detections by AEDGE of the infall, merger and ringdown phases of IMBHs in the ${\cal O}(10^3 - 10^4)$ solar-mass range (darker blue shading). This is similar to the strategy proposed in~\cite{Sesana2016} for the synergistic operation of LISA and LIGO.

$\bullet$ As discussed in~\cite{Carson:2019rda}, combined measurements
by AEDGE and other detectors would provide unparalleled lever arms for probing fundamental
physics by measuring post-Newtonian and post-Minkowskian~\cite{Bern:2019nnu} gravitational parameters, probing Lorentz invariance in GW propagation and the possibility of parity-violating gravity.

In summary, the mid-frequency GW detection capabilities of AEDGE discussed here will play a crucial part  in characterising  the full mass spectrum of black holes and their evolution, thereby casting light on their role in shaping galaxies~\footnote{In addition to this primary astrophysical programme, we note that AEDGE would also be able to measure GWs from galactic white-dwarf (or other) binaries with orbital periods lass than about a minute, a possibility whose interest has been heightened recently by the observation of a binary with orbital period below 7 minutes~\cite{Burdge:2019hgl}.}. \\
~~\\
\mysubsubsection{Cosmological Sources}
~~\\
$\bullet$ Many extensions of the Standard Model (SM) of particle physics
predict first-order phase transitions in the early Universe. Examples include extended electroweak
sectors, effective field theories with higher-dimensional operators and hidden-sector interactions. Extended electroweak models have
attracted particular interest by providing options for electroweak baryogenesis and magnetogenesis: see, e.g., \cite{Ellis:2019tjf}, and offer opportunities for correlating cosmological observables with signatures at particle colliders~\cite{Ellis:2018mja,Ellis:2019oqb}.

\begin{figure}[t]
\centering
\includegraphics[width=0.45\textwidth]{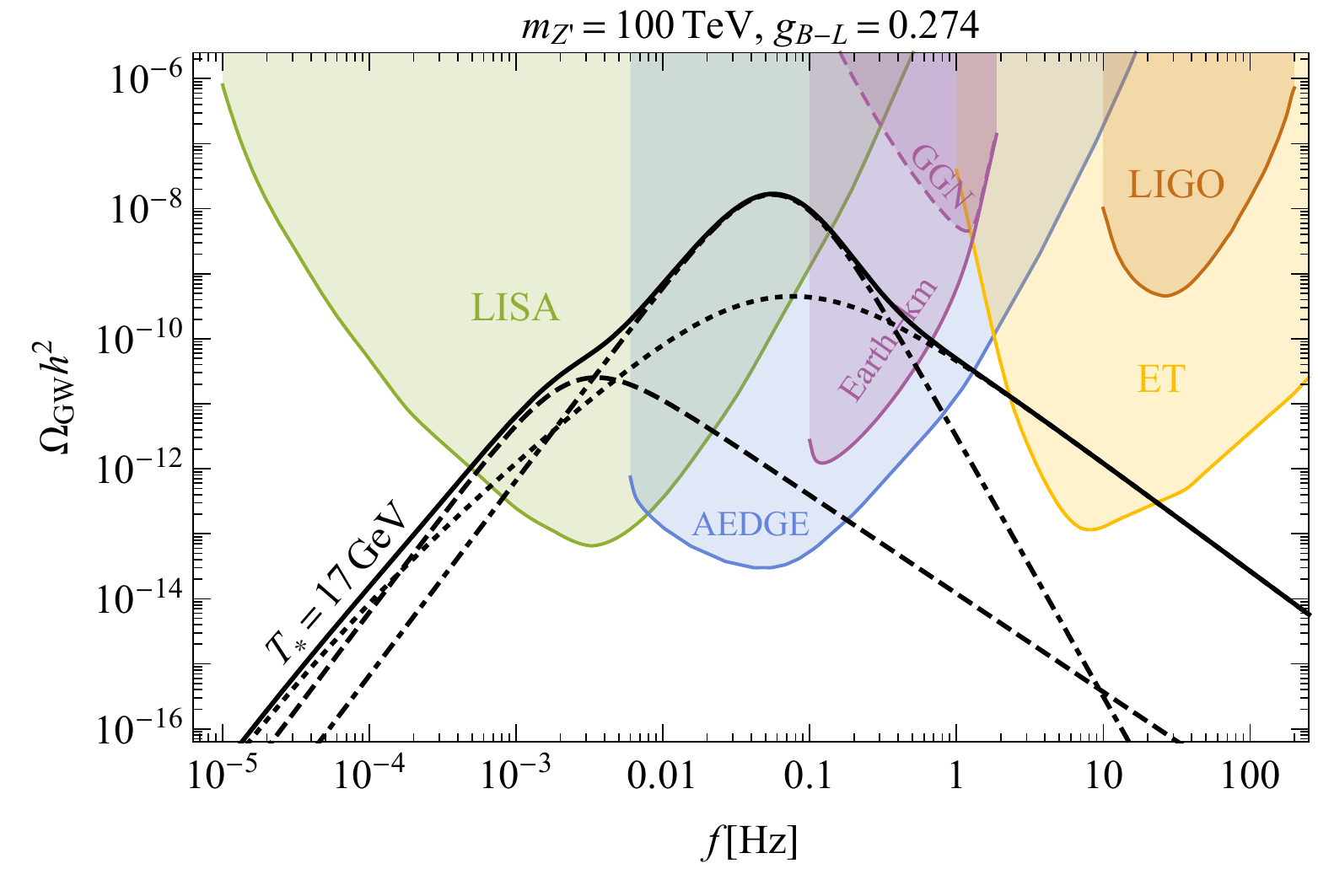}
\hspace{8mm}
\includegraphics[width=0.45\textwidth]{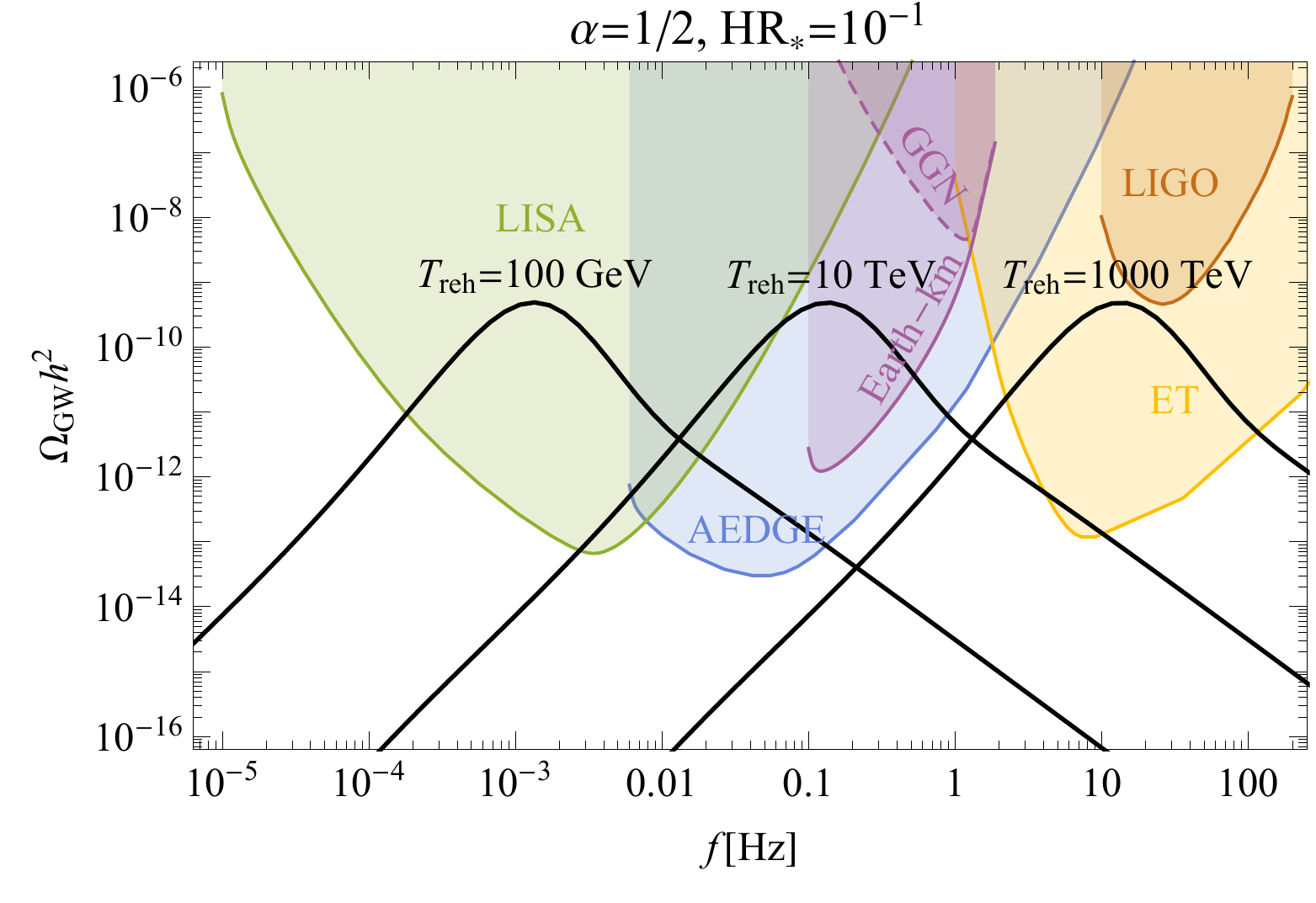}
\vspace{-2mm}
\caption{\it Left panel: Example of the GW spectrum in a classical scale-invariant extension of the SM with a massive $Z^\prime$ boson,
compared with various experimental sensitivities. The dashed line shows the contribution to the spectrum sourced by bubble collisions, the dot-dashed line shows the contribution from sound waves, and the dotted line shows the contribution from turbulence.
Right panel: Examples of spectra with some other reheating temperatures after the transition that may be realized in the same model.
}
\label{staroplot3}
\end{figure}

The left panel of Fig.~\ref{staroplot3} shows one
example of the GW spectrum calculated in a classically scale-invariant extension of the SM with a massive $Z^\prime$ boson, 
including both bubble collisions and the primordial plasma-related sources~\cite{Ellis:2019oqb}. These contributions yield a broad spectrum whose shape can be probed only by a combination 
of LISA and a mid-frequency experiment such as AEDGE, { which is assumed here to be operated at a set of ${\cal O}(10)$ resonant frequencies, whose combined data would yield the indicated sensitivity to a broad spectrum}. A crucial feature
in any model for a first-order phase transition in the early universe
is the temperature, $T_*$, at which bubbles of the new vacuum percolate.
For the model parameters used in the left panel of Fig.~\ref{staroplot3},
$T_* = 17$~GeV. {The GW spectra for parameter choices yielding various values of the reheating temperature, $T_{\rm reh}$, which are typically ${\cal O}(m_{Z^\prime})$ in this model, are shown in the right panel of Fig.~\ref{staroplot3}}. We see that AEDGE would play a key role, fixing the parameters of this classically scale-invariant extension of the SM.

\begin{figure}[t]
\centering
\includegraphics[width=0.5\textwidth]{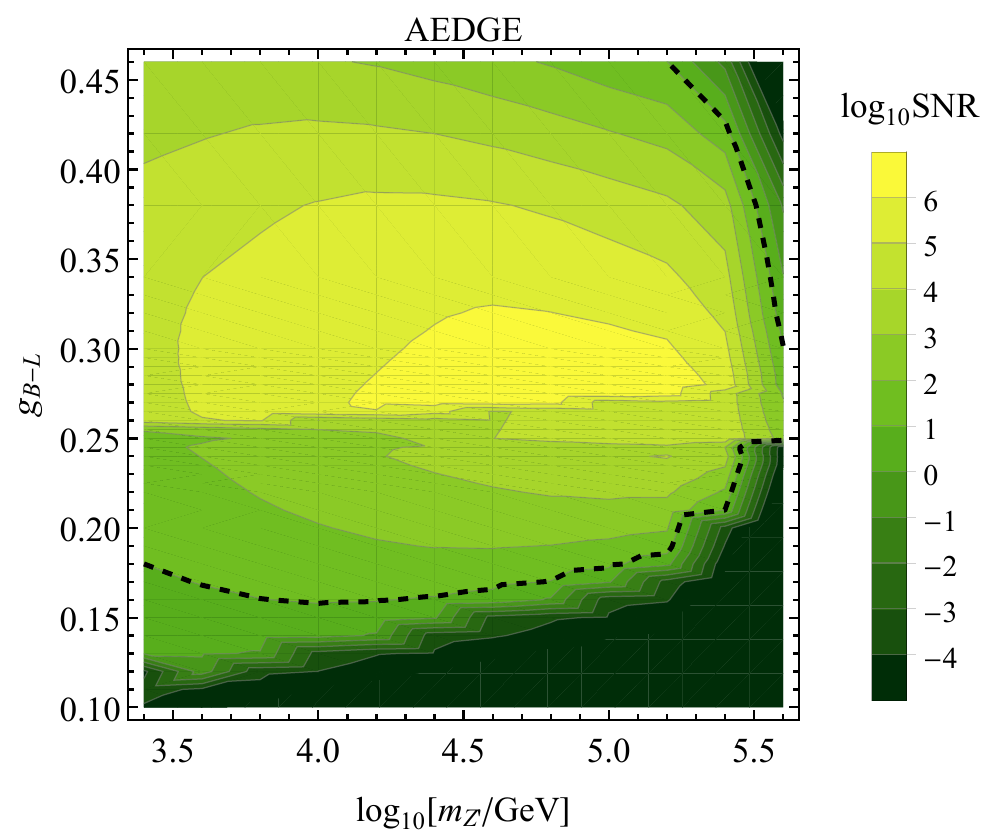}
\caption{\it Signal-to-noise ratio (SNR) achievable with AEDGE in the parameter plane of the classically scale-invariant extension of the SM with a massive $Z^\prime$ boson. The dashed line is the SNR~$=10$ contour.}
\label{staroplot33}
\end{figure}

Fig.~\ref{staroplot33} shows the discovery sensitivity of AEDGE in the parameter space of the classically scale-invariant extension of the SM with a massive $Z^\prime$ boson. We see that AEDGE could measure a signal from a strong phase transition with high signal-to-noise ratio (SNR) all the way down to the present lower limit of a few TeV on the $Z^\prime$ mass from experiments at the LHC, and covering the mass range where such a boson could be discovered at a future circular collider~\cite{Abada:2019lih}. {The SNR is calculated assuming five years of observation time divided between 10 resonance frequencies, whose data are combined.}

%\begin{wrapfigure}[13]{l}{0.5\textwidth}
%\begin{figure}[h!]
%\centering
%\vspace{-0.5cm}
%\includegraphics[scale=0.3]{Marekstring.png}
%\caption{\it Examples of GW spectra from cosmic strings with tension $G \mu =10^{-11}$, for different %values of the maximum  temperature $T_F$
%at which string loops formed.}
%\label{staroplot}
%\end{wrapfigure}

%Other possible cosmological sources of GW signals are cosmic strings. As seen in Fig.~4, these typically give a very broad frequency
%spectrum stretching across the ranges to which LIGO/ET, AION and LISA are sensitive. Since the frequency spectrum is relatively featureless,
%combining measurements of experiments in different frequency ranges will be crucial for pinning down the interpretation of any signal, e.g., the
%maximum  temperature $T_F$ at which string loops formed.\\
%~~\\

\begin{figure}
%\begin{figure}[h!]
\centering
%\vspace{-0.5cm}
\includegraphics[width=0.475\textwidth]{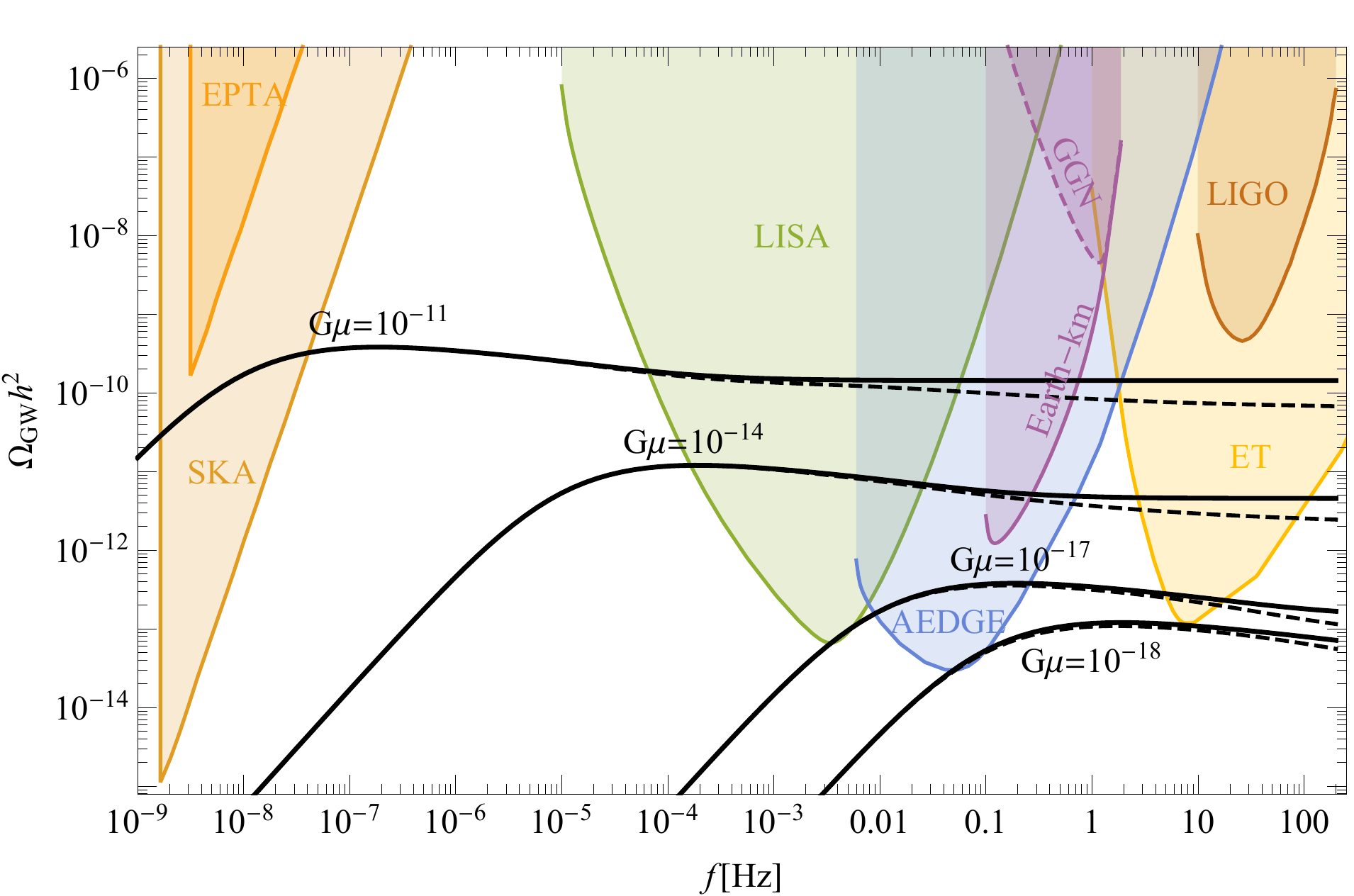}
\includegraphics[width=0.5\textwidth]{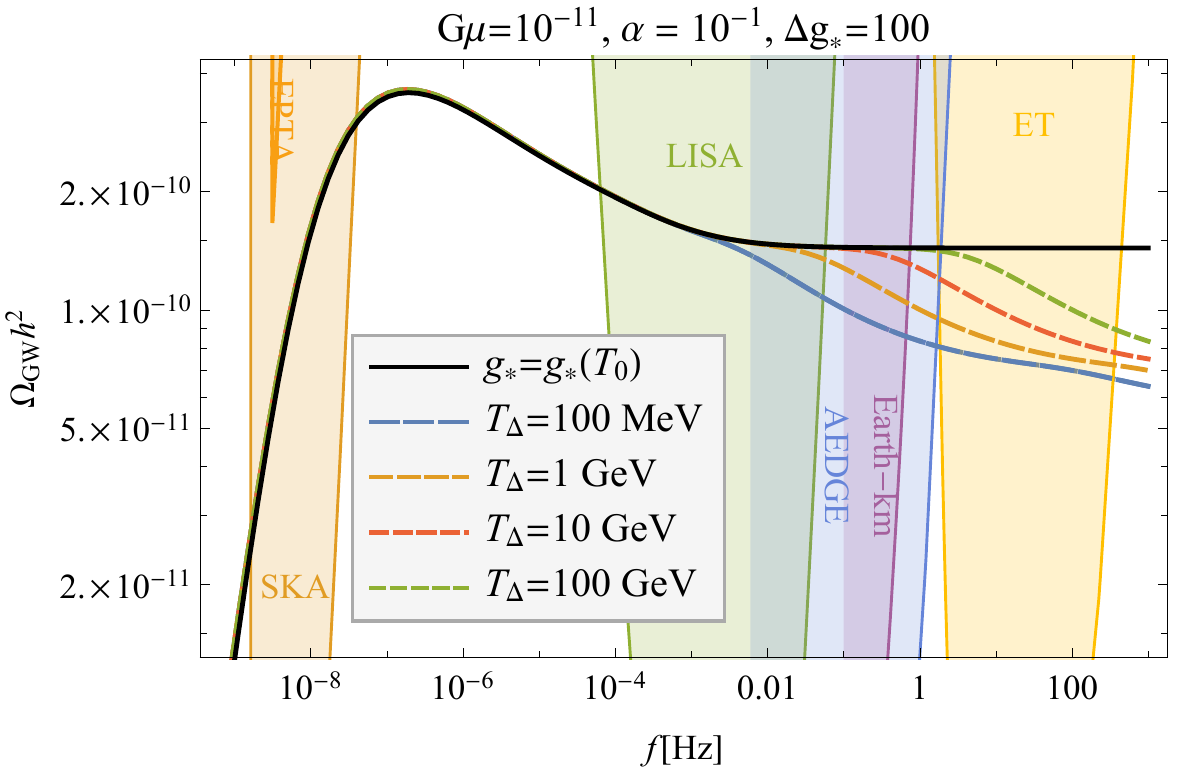}
\caption{\it Left panel: Examples of GW spectra from cosmic strings with differing tensions $G\mu$. The dashed lines show the impact of the variation in the number of SM degrees of freedom. Right panel: Detail of the effect on the GW spectrum for the
case $G \mu = 10^{-11}$ of a new particle threshold at various energies $T_\Delta \ge 100$\,MeV with an increase $\Delta g_* = 100$ in the number of relativistic degrees of freedom.}
\label{staroplot4}
\end{figure}

$\bullet$ Other possible cosmological sources of GW signals include cosmic strings. As seen in the left panel of Fig.~\ref{staroplot4}, these typically give a very broad frequency
spectrum stretching across the ranges to which the LIGO/ET, AEDGE, LISA and SKA~\cite{Bacon:2018dui} experiments are sensitive. {The current upper limit on the string
tension $G \mu$ is set by pulsar timing array (PTA) measurements at low frequencies~\cite{vanHaasteren:2011ni}. LISA will be sensitive to $G\mu=10^{-17}$, while AEDGE and ET could further improve on this sensitivity by an order of magnitude.}
This panel also shows (dashed lines) the impact of including the change in the number of degrees of freedom predicted in the SM. It is apparent that
detailed measurements in different frequency ranges could probe both SM processes such as the QCD phase transition and BSM scenarios predicting new degrees of freedom, e.g., in a hidden sector, or even more significant cosmological modifications such as early matter domination, which would leave distinguishable features in the GW background. This point is illustrated in the right panel of Fig.~\ref{staroplot4}, where we see the effect on the string GW spectrum of a new particle threshold at energies $T_\Delta \ge 100$~MeV with an increase $\Delta g_* = 100$ in the number of relativistic degrees of freedom. Comparing the string GW strengths at different frequencies at the 1 \% level would be sensitive to $\Delta g_* = 2$.

\begin{figure}
%\begin{figure}[h!]
\centering
%\vspace{-0.5cm}
\includegraphics[width=0.6\textwidth]{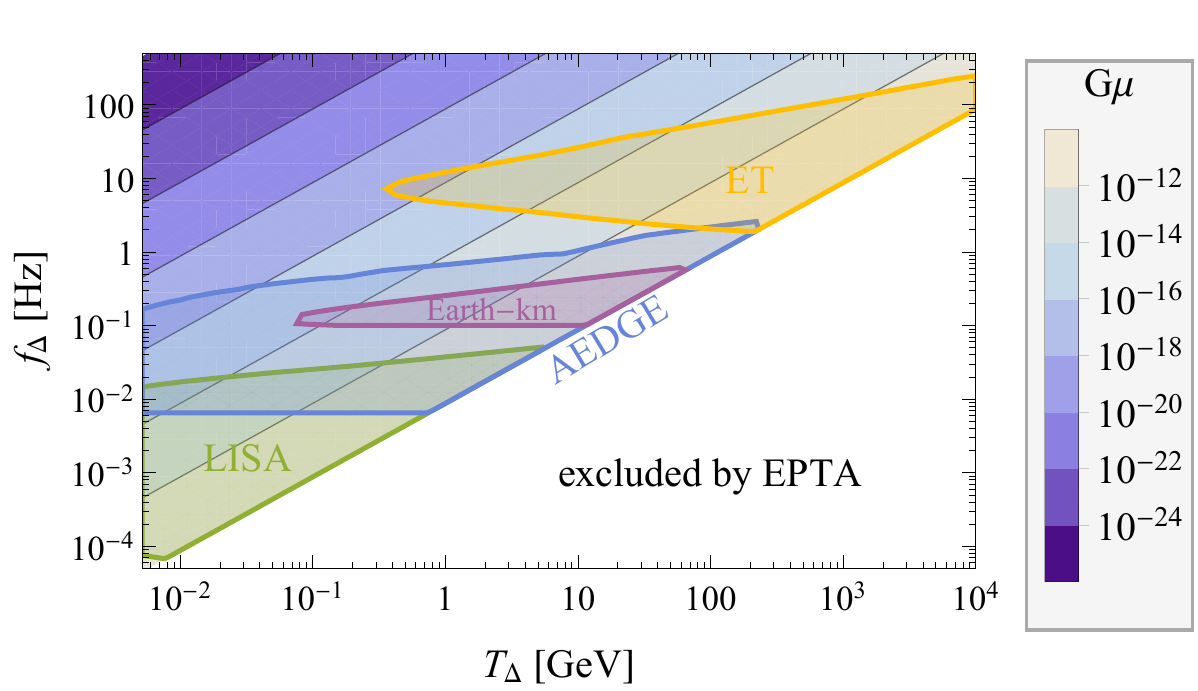}
\caption{\it Frequency $f_{\Delta}$ at which features in the cosmic string GW spectrum appear corresponding to events in the early universe occurring at the indicated temperature $T_{\Delta}$. The shading contours indicate $G\mu$ values of the cosmic string network, and the reach of different experiments are indicated by the coloured regions.}
\label{staroplot5}
\end{figure}

In Fig.~\ref{staroplot5} we show the frequencies at which features would appear in the cosmic string GW spectrum corresponding to events in the early universe occurring at different temperatures. We see that AEDGE would be sensitive in a different range of parameters from ET and LISA.
Figs.~\ref{staroplot4} and \ref{staroplot5} illustrate that probing the plateau in a wide range of frequencies can provide a significant amount of information not only on strings themselves but also on the early evolution of the universe~\cite{Cui:2018rwi}.

\subsection{Other Fundamental Physics}
\label{FP}

Ultra-high-precision atom interferometry has been shown to be sensitive to other aspects of fundamental physics beyond dark matter and GWs, though studies of some such possibilities are still at exploratory stages.  Examples include:\\
$\bullet$ High-precision measurements of the {\it gravitational redshift} and quantum probes of the {\it equivalence principle}~\cite{Tino2019};\\
$\bullet$ The possibility of {\it detecting astrophysical neutrinos} that traverse the Earth with high fluxes though small cross-section: see, e.g., \cite{Becker:2007sv}. The great advantage of interferometers in this case is that they are sensitive to very small or even vanishing momentum transfer. Whilst current sensitivities seem far from accessing any interesting background \cite{Alonso:2018dxy}, the analyses of this possibility have not been comprehensive;\\
$\bullet$ {\it Probes of long-range fifth forces}: Since atom interferometry can be used to detect the gravitational field of Earth \cite{Peters_2001}, a set up with interferometers at different heights seems a natural one to study the possibility of any other long-range fifth force that couples to matter in ways different from gravity. The search for long-range forces is a very active area of research beyond the SM, with natural connections to dark matter and modified gravity, see, e.g.,  \cite{Safronova:2017xyt}, and universally-coupled Yukawa-type fifth forces over these scales are already well constrained by classical searches for fifth forces~\cite{Dimopoulos:2008hx}; \\
$\bullet$ {\ Tests of general relativity}: A set-up with atom interferometers at different values of the gravitational potential also facilitates
measurements of higher-order general-relativistic corrections to the gravitational potential around the Earth. The leading higher-order effects are due to the gradient of the potential, and corrections due to the finite speed of light, and D{\" o}ppler shift corrections to the photon frequency;\\
$\bullet$ {\it Constraining possible variations in fundamental constants}: A comparison of interferometers at different time and space positions may be useful to test possible variations of fundamental constants in these two domains. There are different motivations for these searches that can be found in \cite{Uzan:2002vq,Martins:2019qxe};\\
$\bullet$ {\it Probing dark energy}: The main driver of current cosmological evolution is a puzzling substance that causes the acceleration of the expansion of space-time. This `dark energy' is supposed to be present locally and one can try to use precise experiments to look for its local effects. This possibility comes in at least two flavours. One can argue that dark energy models naturally involve dynamical ultra-light fields. If the SM is coupled to them, the fundamental properties of nature would be time- and space-dependent. Another possibility comes from specific models where the dark energy candidate modifies the laws of gravity, where atom interferometry experiments have proved a particularly powerful technique for constraining popular models \cite{Jaffe:2016fsh,Sabulsky:2018jma};\\
$\bullet$ {\it Probes of basic physical principles}. These include probing Bell inequalities and testing the foundations of quantum mechanics and Lorentz invariance. It has been suggested that some ideas beyond the standard postulates of quantum mechanics (for instance linearity and collapse models) may be tested with precise interferometry of quantum states, see, e.g., \cite{Ellis:1983jz,Banks:1983by,Ghirardi:1985mt,Weinberg:2016uml}, and atom interferometers have been proposed as test of Lorentz invariance and gravitation in \cite{Chung:2009rm}.\\

%The previous points illustrate that atomic interferometry is a promising direction towards testing different fundamental aspects of Nature. This proposal aims to clarify how AION can play a role in probing such possibilities. At this stage one can already realize that AION brings new configurations for the interferometers. The aim of the AION proposal is also to generate a community with a continous dialogue between atomic physicist and particle physicists and astrophysicsts essential to study the previous points in the long run.

%\section{Scientific Requirements}

\section{Experimental Considerations}\label{MC}

In this Section we describe a conceptual detector design that can accomplish the science goals  outlined in this document. This basic design requires two satellites operating along a single line-of-sight and separated by a long distance. The payload of each satellite will consist of cold atom technology as developed for state-of-the-art atom interferometry and atomic clocks. For the science projections presented here, we assume a minimum data-taking time of 3 years, which requires a mission duration of at least 5 years, while 10 years would be an ultimate goal. 

%{\it As two satellites are essential to accomplish the science goals, AEDGE has to be an L mission.}          
{\it As two satellites are needed to accomplish its science goals, the AEDGE mission planning costs are estimated to be in the range of an L-class mission. However, in view of the international interest in the AEDGE science goals, the possibility of international cooperation and co-funding of the mission may be investigated.} 

\subsection{Representative Technical Concept}

As we discuss in Section~\ref{TR}, there are several cold atom projects based on various technologies that are currently under construction, planned or proposed, which address the principal technical challenges and could be considered in a detailed design for a mission proposal and corresponding satellite payload. However, all of these options require the same basic detector and mission configuration outlined above. For the option presented in this White Paper we have chosen to base our discussion on the concept outlined in in~\cite{Graham:2012sy,Graham:2016plp,Graham:2017pmn,Tino2019}, which is currently the most advanced design for a space mission.   

\begin{figure}[t]
\centering
\includegraphics[width=0.5\textwidth]{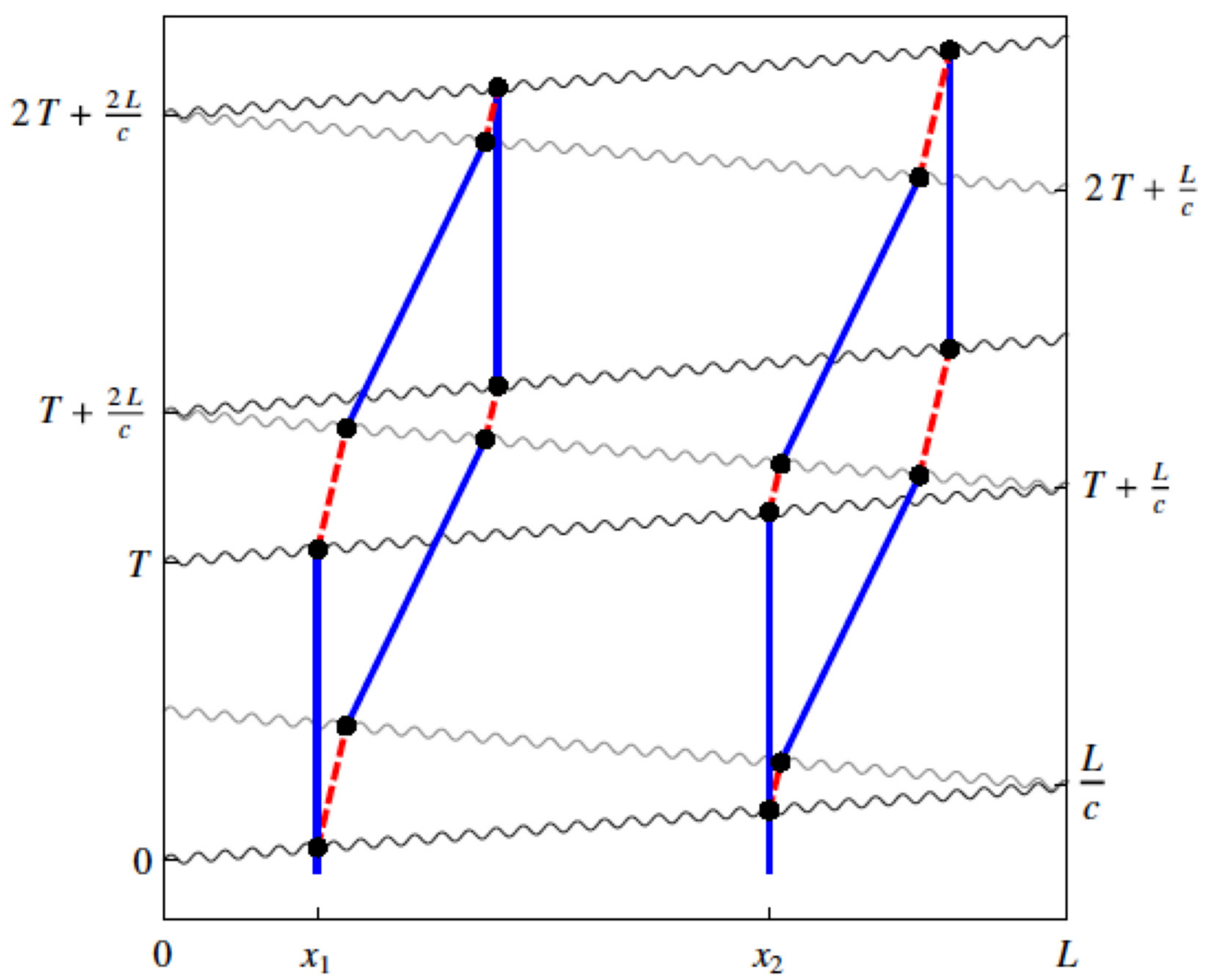}
\caption{\it Space-time diagram of the operation of a pair of cold-atom interferometers based on single-photon transitions between the ground state (blue) and the excited state (red dashed). The laser pulses (wavy lines) travelling across the baseline from opposite sides are used to divide, redirect, and recombine the atomic de Broglie waves, yielding interference patterns that are sensitive to the modulation of the light travel time caused by DM or GWs (from~\cite{Graham:2012sy}). For clarity, the sizes of the atom interferometers are shown on an exaggerated scale.}
\label{space-time}
\end{figure}

This concept links clouds of cold atomic strontium in a pair of satellites in medium earth orbit (MEO) via pulsed continuous-wave lasers that induce the 698\,nm atomic clock transition, and detect momentum transfers from the electromagnetic field to the strontium atoms, which act as test masses in the double atom interferometer scheme illustrated in Fig.~\ref{space-time}. The lasers are separated by a large distance $L$, the paths of the light pulses are shown as wavy lines, and the atom interferometers, which are represented by the two diamond-shaped loops on an enlarged scale, are operated near them. Laser pulses transfer momenta $\hbar k$ to the atoms and toggle them between the ground state and the excited state.  Thus they act as beam splitters and mirrors for the atomic de Broglie waves, generating a quantum superposition of two paths and then recombining them. As in an atomic clock, the phase shift recorded by each atom interferometer depends on the time spent in the excited state, which is related directly to the light travel time across the baseline, namely $L/c$.

A single interferometer of the type described here, e.g., the interferometer at position $x_1$ in Fig.~\ref{space-time}, would be sensitive to laser noise, 
but a crucial experiment has demonstrated~\cite{Hu:2017kp} that this can be substantially suppressed by the differential measurement between the two interferometers at $x_1$ and $x_2$ as suggested in~\cite{Graham:2012sy}. The sensitivity of a single such interferometer could be substantially improved in the two-interferometer configuration outlined here by measuring the differential phase shift between the widely-separated interferometers~\cite{Graham:2012sy}. The GW (or DM) signal provided by the differential phase shift is proportional to the distance $L$ between the interferometers, whereas the laser frequency noise largely cancels in the differential signal. 

\begin{figure}[h!]
\vspace{0.5cm}
\centering
\includegraphics[width=0.95\textwidth]{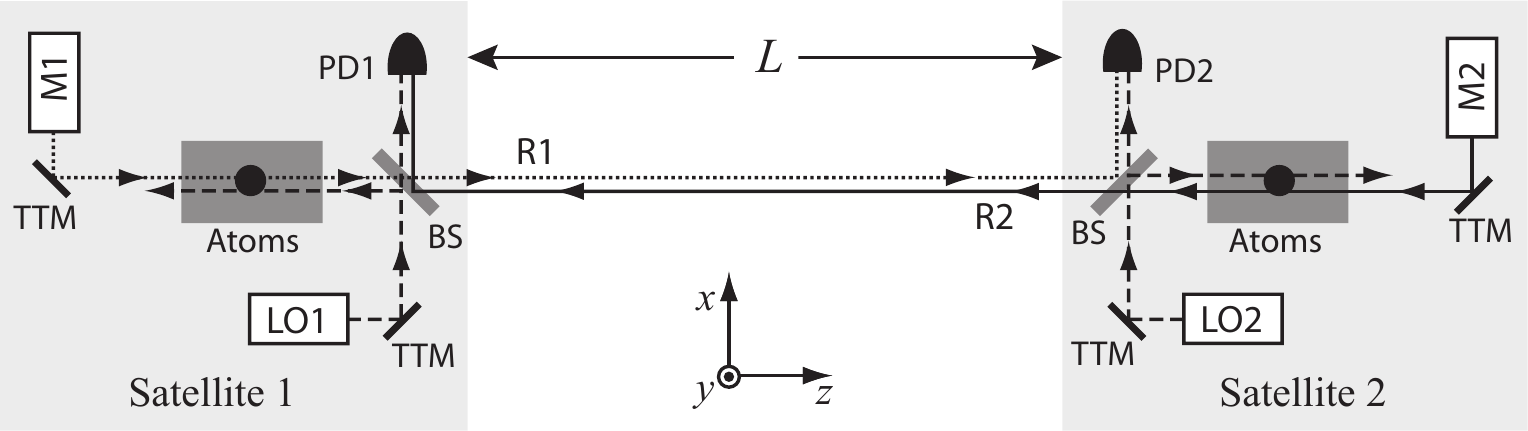}
\caption{\it Possible experimental scheme.  The beams of the two master lasers M1 and M2 are shown as dotted and solid lines, respectively, together with the corresponding reference beams  between the satellites, R1 and R2.   Two local oscillator lasers LO1 and LO2 (dashed  lines) are phase-locked with R2 and R1, respectively. Photodetectors PD1 and PD2 measure the heterodyne beatnote between the reference beams R2 and R1 and the corresponding local lasers LO1 and LO2, respectively, providing feedback for the laser link. Non-polarizing beam splitters are denoted by BS, and tip-tilt mirrors used for controlling the directions of the laser beams are denoted by TTM. For clarity, small offsets between overlapping laser beams have been introduced. Figure taken from~\cite{Graham:2017pmn}.}
\label{sat-concept}
\end{figure}

Based on this approach using two cold-atom interferometers that perform a relative measurement of differential phase shift, we propose a mission profile using a pair of satellites similar to that used for atomic gravity gradiometers~\cite{Snadden:1998zz,Sorrentino:2013uza}, which is shown in Fig.~\ref{sat-concept}. As the atoms serve as precision laser frequency references, only two satellites operating along a single line-of-sight are required to sense gravitational waves. The satellites both contain atom interferometers that are connected by laser pulses propagating along  the positive and negative $z$ directions in the diagram, and the clouds of ultracold atoms at the ends of the baseline of length $L$ act as inertial test masses. There are intense master lasers (M1 and M2) in the satellites,  which drive the atomic transitions in the local atom interferometers.  After interaction with the atoms, each master laser beam is transmitted by the beam splitter (BS) out of the satellite,  and  propagates towards the other satellite, and R1 and R2 are beams from satellite 1 and 2, respectively, that play the roles of reference beams.

Intense local lasers LO1 and LO2 are used to operate the atom interferometers at each end of the baseline. These otherwise independent local lasers are connected by reference laser beams R1 and R2 that are transmitted between the two spacecraft, and the phases of the local lasers are locked/monitored with respect to the incoming wavefronts of these reference lasers, as illustrated in Fig.~\ref{sat-concept}.  A detailed description is available in~\cite{Hogan:2015xla,Graham:2016plp,Graham:2017pmn}.

In addition to photodetectors PD1 and PD2 for measuring the phase differences between the two beams in both satellites, the spatial interference patterns are characterized by quadrant detectors (or cameras), enabling the pointing directions and spatial modes of the two lasers to be well matched using appropriate feedback.  Feedback applied to the tip-tilt mirrors (TTMs) in Fig.~\ref{sat-concept} can then be used to control the angles of the local lasers. Similarly, the angle of the master laser itself can be controlled by comparing it to the local laser direction and using another TTM. 

With satellites in MEO, the measurement baseline re-orients on a time scale that is short compared to the expected duration of the GW signals from many anticipated sources.  This allows efficient determination of the sky position and can provide polarization information.  The relatively short measurement baseline, compared to LISA, provides good sensitivity in the 0.01\,Hz to 1\,Hz frequency band, intermediate between the LISA and LIGO antenna responses, and suited to GW astronomy, cosmology and DM searches, as described above. 

\subsection{Sensitivity Projections}
\begin{table}[t]
  \begin{center}
    \caption{List of basic parameters of strontium atom interferometer designs for AEDGE and a benchmark 1-km terrestrial experiment using similar technologies: length of the detector L; interrogation time of the atom interferometer $T_{\rm int}$; phase noise $\delta\phi_{\rm noise}$; and the total number of pulses $n_p^{\rm max}$, where $n$ is the large momentum transfer (LMT) enhancement and $Q$ the resonant enhancement. The choices of these parameters predominately define the sensitivity of the projection scenarios\cite{Graham:2017pmn}.}
    \vspace{5mm}
    \label{tab:parameters}
    \begin{tabular}{c|c|c|c|c} 
    \textbf{Sensitivity} & $L$ & $T_{\mathrm{int}}$ &  $\delta\phi_{\rm noise}$ & $n_p^{\rm max}=2Q(2n-1)+1$ \\
      \textbf{Scenario} & [m] & [sec] & [$1/\sqrt{{\rm Hz}}$] &  [number]\\
      \hline
      Earth-km & 2000 & 5 & $0.3 \times 10^{-5}$ & 40000\\
      AEDGE & $4.4\times10^{7}$  & 300 & $10^{-5}$ & 1000\\
    \end{tabular}
  \end{center}
  \vspace{-2mm}
\end{table}
In order to establish sensitivity estimates for the different physics goals described above, we have to choose a concrete scenario and define quantitative projections. 

For example, a GW would modify the light travel time across the baseline of the two-satellite system, varying the time spent in the excited state by atoms at each end of the baseline, generating a differential phase shift between the two atom interferometers.  The phase response of the detector can be written as $\Delta\Phi_\text{grad}(t_0)=\Delta\phi \cos{(\omega t_0+\phi_0)}$, where $\omega t_0+\phi_0$ is the phase of the GW at time $t_0$ at the start of the pulse sequence.  The resulting amplitude of the detector response is~\cite{Graham:2016plp}:
\begin{equation}
\Delta\phi= k_\text{eff} h L \frac{\sin\!{(\omega Q T)} }{\cos\!{(\omega T/2)}}\mathrm{sinc}{\big(\tfrac{\omega n L}{2 c}\big)}  \sin\!{\big(\tfrac{\omega T}{2}\!-\!\tfrac{\omega(n-1)L}{2c}\big)} \, ,
\label{Eq:PhaseShift}
\end{equation}
where $\hbar k_\text{eff}$ is the effective momentum transfer, and $k_\text{eff}\equiv n \omega_A/c$ for an $n$-pulse sequence generating an atomic transition with level spacing $\hbar \omega_A$.  The response (\ref{Eq:PhaseShift}) is peaked at the resonance frequency 
$\omega_r \equiv \pi/T$ and exhibits a bandwidth $\sim\!\omega_r/Q$.  The amplitude of the peak phase shift on resonance is
\begin{equation}
\Delta\phi_\text{res}=2Q k_\text{eff} h L\,\mathrm{sinc}{\big(\tfrac{\omega_r n L}{2 c}\big)} \cos\!{\big(\tfrac{\omega_r (n-1) L}{2 c}\big)} \, ,
\label{Eq:PeakPhaseShift}
\end{equation}
which reduces in the low-frequency limit $\omega_r \ll \tfrac{c}{n L}$  to $\Delta\phi_\text{res}\approx 2 Q k_\text{eff} h L$.  The phase response shows an $n$-fold sensitivity enhancement from large momentum transfer (LMT). The interferometer can be switched from broadband to resonant mode by changing the pulse sequence used to operate the device (changing $Q$)~\cite{Graham:2016plp}, resulting in a $Q$-fold enhancement. 

For the sensitivity projections of AEDGE presented in this paper we assume that operation is performed mainly in the resonant mode, while also providing estimates for broadband operation for comparison. In order to generate the sensitivity curve for, e.g., a GW signal, from the phase response, we calculate the minimum strain $h$ that is detectable given a phase noise spectral density $\delta\phi_{\rm noise}$. We optimize the LMT enhancement $n$ for each frequency and resonant enhancement $Q$, taking into account the detector design constraints, which include the limits on the total number of pulses, $n_p^{\rm max} = 2Q(2n-1)+1$, and on the maximum interferometer duration, $2TQ < T_{\rm int}$, where $T_{\rm int}$ is the time over which the atom interferometer is interrogated. 
Furthermore, as we assume in the design outlined above that the interrogation region of the atoms is placed within the satellite, the wavepacket separation  $\Delta x=\hbar k_\text{eff} (T/m)$, where $m$ is the atom mass, is constrained to be less than 90 cm. As discussed in~\cite{Graham:2016plp, Graham:2017pmn}, this constraint limits the amount of LMT enhancement. Using resonant enhancement while reducing LMT allows the interferometer region to remain small, but it has an impact on the achievable sensitivity when setup is operated in broadband mode. An alternative design places the interrogation region outside the satellite~\cite{Dimopoulos:2008sv}. This setup would support LMT values closer to what can be achieved in ground-based setups, which would not only increase broadband sensitivity but also make it possible to probe even lower frequencies. However, operating the interferometers in space would incur additional technical challenges such as vacuum stability, solar radiation shielding and magnetic field effects. While these challenges seem surmountable, conservatively we focus our sensitivity  projections here on a design in which the atom interrogation region is within the satellite, which requires resonant mode operation to achieve maximal sensitivity. In the future, further investigations of using a much larger interrogation region in space could change this design choice. 

 This resonant mode strategy provides significant sensitivity to a stochastic background of gravitational waves,
 {  e.g., of cosmological origin. To indicate the sensitivity estimates for the density of GW energy, $\Omega_\text{GW}$, we use power-law  integration~\cite{Thrane:2013oya} to display an envelope of power-law signals for each given frequency detectable with an assumed ${\rm SNR}=10\, $.  In the calculation for AEDGE we assume five years of observation time divided between 10 logarithmically-distributed resonance frequencies and sum the signal from the total running time of the experiment.}
  {We have verified that changing this scanning strategy by using a different number of resonant frequencies does not have a strong impact on the resulting sensitivity.}
%  Following the discussion in~\cite{Graham:2016plp} we estimate the $95 \%$ confidence limit on such a stochastic background as
% \begin{equation}
% \label{eqn: stochastic sensitivity}
% \Omega_\text{GW} (f) = \frac{\pi c^2 f^3}{\rho_c G | \gamma \left( \vec{x}_1, \vec{x}_2, f \right) |} \sqrt{\frac{2}{\tau_\text{int} \Delta f}} (1.645) h_n^2 (f) \, ,
% \end{equation}
% where $\tau_\text{int}$ is the total averaging time of the experiment, $\gamma$ is a geometric factor that takes into account the positions of the two detectors (which we take equal to its maximum value $\frac{8 \pi}{5}$, though it will probably be slightly smaller in a real configuration), and $\rho_c$ is the critical density of the Universe.  Here $h_n$ is the amplitude spectral density of the strain noise in the gravitational wave detector. For the bandwidth we take $\Delta f \sim \frac{f_r}{Q}$, where $f_r=\omega_r/2\pi$ is the resonant frequency of the detector.  The sensitivity that can be reached at each frequency operating in resonant mode is than calculated for an integration time of $\tau_\text{int} = 1~\text{year}$.
%  For sensitivity estimates of $\Omega_\text{GW}$ in broadband mode we use power law  integration~\cite{Thrane:2013oya} showing an envelope of power-law signals for each given frequency detectable with assumed ${\rm SNR}=10$. 
  {These curves thus have the property that any power-law signal touching them would give the required SNR in the indicated experiment. For ease of comparison, we also assumed five years of operation for each of the other experiments shown.} 
 
 %including terrestrial km version of AEDGE, however, not for the satellite version of AEDGE optimised to run in the resonant mode as explained above.

The quantitative projections for the DM and GW signals we presented in the previous Sections are based on the following scenarios:

\begin{itemize}
    \item {\bf Earth-km}: This scenario represents the sensitivity estimate of a terrestrial detector at the km-scale using typical parameters that are projected to be achieved in the future. This sets the benchmark for comparison with the space-based AEDGE.
    \item {\bf AEDGE}: This scenario represents the sensitivity estimate of a space-based detector using parameters that could be achieved for this set-up. This sets the benchmark for the sensitivity of space-based detector proposed in this White Paper.
\end{itemize}

The values of the basic parameters assumed for the different sensitivity scenarios are listed in Table~\ref{tab:parameters}. These parameters dominate in determining the sensitivities for the DM and GW projections presented in Sections~\ref{ULDM} and \ref{GW}, respectively.   

\section{Technological Readiness}
\label{TR}

AEDGE will benefit from the experience gained with LISA Pathfinder in free-fall control and LISA itself in operating laser interferometers over large distances. We have identified the following three additional high-level technical requirements that are critical for AEDGE: 
\begin{itemize}
    \item Demonstrate reliable functioning of atom interferometry on a large terrestrial scale $\gtrsim 100$m; 
     \item Demonstrate that the design parameters assumed here, such as the LMT enhancement, phase noise control, interrogation time, etc., can be achieved;  
     \item Demonstrate the robustness of cold atom technology in the space environment.  
\end{itemize}
%\subsection{Road map to Technical Readiness} 
Several terrestrial atom interferometer projects that would serve as demonstrators for different technologies are under construction, planned or proposed, representing a qualitative change in the state of technological readiness since the SAGE project~\cite{Tino2019} was reviewed by ESA in 2016. As described below, they should be able to show how the above-mentioned technical requirements can be met and demonstrate TRL5 technology readiness (according to ISO Standard 16290).

      $\bullet$  Three large-scale prototype projects at the 100-m scale are funded and currently under construction, namely MAGIS-100 in the US, MIGA in France, and ZAIGA in China. These will demonstrate that atom interferometry at the large scale is possible, paving the way for terrestrial km-scale experiments. Assuming that large-scale prototyping is successful within five years, extending the technology to the km scale will be the next step. There are projects to build one or several more km-scale detectors in the US (at the Sanford Underground Research facility, SURF), in Europe (MAGIA-advanced, ELGAR) and in China (advanced ZAIGA) that would serve as the ultimate technology readiness demonstrators for AEDGE. It is foreseen that by about 2035 one or more km-scale detectors will have entered operation. 
     
      $\bullet$  In parallel to these large-scale prototype projects, several other cold atom projects are in progress or planned, demonstrating the general readiness of the technology including the scaling of the basic parameters that are required for AEDGE. In fact, the basic requirements for AEDGE in terms of atom interferometry are more relaxed than those one of the km-scale terrestrial detectors, as the main sensitivity driver for AEDGE will be the long baseline, and its requirements for the basic parameters of atom interferometry are less stringent than in the km-scale projects.      
     
     $\bullet$  Several cold atom experiments (CACES~\cite{Liu:2018kn}, MAIUS~\cite{MAIUS}, CAL~\cite{CAL}) and underlying optical key technologies (FOKUS~\cite{Lezius}, KALEXUS~\cite{Dinkelaker}, JOKARUS~\cite{PhysRevApplied.11.054068}) have already demonstrated reliable operation in space, and much more experience will be gained in the coming years. 
\\

We now summarize the statuses of some of the key ``AEDGE pathfinder" experiments:

$\bullet$ The Matter-wave laser Interferometric Gravitation Antenna (MIGA)
experiment~\cite{Canuel:2017rrp}, a double 150-m-long optical cavity in Rustrel, France is
fully funded and currently in the final phase of construction.
MIGA aims at demonstrating precision measurements of gravity with cold atom sensors in a large-scale instrument and at studying associated applications in geoscience and fundamental physics. {MIGA will employ an array of atom interferometers along the same optical link to mitigate the main noise contribution at low frequency represented on Earth by Newtonian noise~\cite{Chaibi:2016dze}.}
In particular, it will assess future potential applications of atom interferometry to gravitational wave detection in the mid-frequency band between $\sim 0.1$ and 10~Hz intermediate between LISA and LIGO/Virgo/KAGRA/INDIGO/ET/CE. 

$\bullet$ The MAGIS project~\cite{Graham:2017pmn} in the US plans a series of interferometers using cold atoms with
progressivly increasing baselines of $\sim 10$m, $\sim 100$m, and $\sim 1$km. The first step 
is funded and under construction at Stanford, the second step is also funded and being prepared at Fermilab, 
and the third step is planned for a km-scale vertical shaft at SURF.

$\bullet$ The Zhaoshan long-baseline Atom Interferometer Gravitation Antenna (ZAIGA) is an
underground laser-linked interferometer facility~\cite{Zhan:2019quq} 
under construction near Wuhan, China. It has
an equilateral triangle configuration with two atom interferometers separated by a km in each arm, 
a 300-meter vertical shaft equipped with an atom fountain and atomic clocks, and 1-km-arm-length 
optical clocks linked by locked lasers. It is designed for a comprehensive range of experimental research on gravitation and related problems including GW detection and high-precision tests of the equivalence principle. 

$\bullet$ Building upon the MAGIA experiment~\cite{Rosi:2014kva,MAGIA},
MAGIA-Advanced is an R\&D
project funded by the Italian Ministry for Research and the INFN for a large-scale atom interferometer based on ultracold rubidium and strontium atoms. In addition to laboratory activity, the team is investigating the possibility of a 100-300 m atom interferometer to be installed in a vertical shaft in Sardinia. Its main goals are GW observation and the search for DM.

$\bullet$ ELGAR is a European initiative to build a terrestrial infrastructure based on cold atoms for GW detection with potential applications also for other aspects of gravitation and fundamental physics such as DM. ELGAR will use a large scale array of correlated Atom Interferometers. A White Paper about this infrastructure is being prepared~\cite{Sabulsky2}.

$\bullet$ The AION project in the UK~\cite{AION} proposes a series of atom interferometers
baselines of $\sim 10$m, $\sim 100$m, and $\sim 1$km, similar to MAGIS, with which it will be networked {` a} la LIGO/Virgo. The first stage
would be located in Oxford, with sites for the subsequent stages awaiting more detailed study.\\

The above terrestrial projects will demonstrate various concepts for large-scale cold atom interferometers 
and provide valuable operational experience. In addition there are ongoing NASA, Chinese, ESA, German and French projects to conduct cold atom experiments in space, some of which have already provided operational experience with cold atoms in space or microgravity environments:\\

$\bullet$ NASA recently installed the Cold Atom Laboratory (CAL) experiment on the ISS. It is reported that the CAL system has been performing nominally and that rubidium Bose-Einstein condensates (BECs)
have subsequently been produced in space on nearly a daily basis~\cite{CAL}~\footnote{We will benefit from in-team expertise in numerical calculations of BECs - see~\cite{Balas} and references therein.}, and the continuation of the CAL science
programme will include an atomic interferometer;

$\bullet$ The Chinese Atomic Clock Ensemble in Space (CACES) demonstrated in-orbit operation of an atomic clock based on laser-cooled rubidium atoms~\cite{Liu:2018kn}.

$\bullet$ The Atomic Clock Ensemble in Space (ACES/PHARAO) project led by ESA plans to install ultra-stable atomic cesium clocks on the ISS, enabling several areas of research including tests of general relativity and string theory, and very long baseline interferometry~\cite{ACES,ACES-PHARAO};

$\bullet$ The Bose-Einstein Condensate and Cold Atom Laboratory (BECCAL) is a bilateral project of NASA and the German Aerospace Center (DLR) for a multi-purpose facility on the international space station, based in the heritage of drop-tower (QUANTUS~\cite{Muntinga:2013pta}) and sounding-rocket experiments (MAIUS~\cite{MAIUS}). It will enable a variety of experiments in atom optics and atom interferometery, covering a broad spectrum of research ranging from fundamental physics to studies for applications in earth observation. It is also intended as a pathfinder for future space missions~\cite{BECCAL}.

%$\bullet$ The BECCAL project plans to bring BECs from ground to space,
%building on experience with drop-tower and sounding-rocket experiments (QUANTUS~\cite{Muntinga:2013pta} and MAIUS~\cite{MAIUS}).
%Scientific applications include atom interferometry to search for dark energy/DM
%and test quantum mechanics, as well as quantum information studies. It is intended as %a pathfinder for future space missions~\cite{BECCAL}.\\
$\bullet$ The ICE experiment operates a dual-species atom interferometer in weightlessness in parabolic flights~\cite{Barrett2016}, and recently reported the all-optical formation of a BEC in the microgravity environment obtained on an Einstein elevator~\cite{Condon2019}.

$\bullet$ In the context of the ISS Space Optical Clock (I-SOC) project of ESA~\cite{SOC1,SOC2} to use cold strontium atoms in space to compare and synchronize atomic clocks worldwide (which can also be used to look for topological DM), ESA is running a development programme aimed at increasing the TRL of strontium-related laser technology. Industrial consortia are currently developing 462 nm and 689 nm lasers, a laser frequency stabilization system, a 813 nm lattice laser, an ultrastable reference cavity and a two-way time/frequency microwave link.\\

For completeness, we also mention other proposals for atomic
experiments in space to probe fundamental physics:

$\bullet$ STE-QUEST is a fundamental science mission that was originally proposed for launch within the ESA Cosmic Vision programme, aimed at probing various aspects of Einstein's theory of general relativity and testing the weak equivalence principle. It features a spacecraft with an atomic clock and an atom interferometer~\cite{STE-QUEST}. This mission is also the subject of a Voyage 2050 White Paper.

$\bullet$ Some of the present 
authors proposed the Space Atomic Gravity Explorer (SAGE) 
mission to the European Space Agency in 2016 in response to a Call for “New Ideas”~\cite{Tino2019}, with the scientific objectives to investigate GWs, DM and other fundamental aspects of gravity, as well as the
connection between gravitational physics and quantum physics, combining quantum sensing and quantum communication based on recent impressive advances in quantum technologies for atom interferometers, optical clocks, microwave and optical links.
%As in this project, the cold atoms used in AEDGE will be kept in free fall using the techniques developed for LISA and demonstrated very successfully by the LISA Pathfinder mission.

$\bullet$ The SagnAc interferometer for Gravitational wavE proposal 
(also called SAGE)~\cite{Lacour:2018nws} was envisaged to detect
GWs with frequency $\sim 1$~Hz using multiple CubeSats on ballistic trajectories in geostationary orbit.

$\bullet$ The Atomic Interferometric Gravitational-Wave Space
Observatory (AIGSO) has been proposed in China~\cite{AIGSO}.\\

AEDGE will also benefit from studies for the Search for Anomalous Gravitation using Atomic Sensors (SAGAS) project~\cite{SAGAS} and the past Space Atom Interferometer (SAI) project~\cite{SAI1,SAI2}, and will maintain contacts with CERN, with a view to applying as a recognized experiment when funded.

\section{Summary}
\label{summary}

The nature of DM is one of the most important and pressing in particle physics and cosmology, and one of the favoured possibilities is that it is provided by coherent waves of some ultra-light boson.
As we have illustrated with some specific examples, AEDGE will be able to explore large ranges of the parameter spaces of such models, complementing the capabilities of other experiments.

Experience with electromagnetic waves shows the advantages
of making astronomical observations in a range of different frequencies, and the same is expected to hold in the era of gravitational astronomy.
There are advanced projects to explore the GW spectrum with maximum sensitivities at frequencies $\gtrsim 10$~Hz and below $\lesssim 10^{-2}$~Hz, but no approved project has peak sensitivity in the mid-frequency band between them.
As we have discussed, the mergers of intermediate-mass black holes, first-order phase transitions in the early universe and cosmic strings are among the possible GW sources that could produce signals in the mid-frequency band.
As we have also discussed, AEDGE would be ideal for exploiting these scientific opportunities, complementing other experiments and offering synergies with them.

Other possible opportunities for AEDGE in fundamental physics, astrophysics and cosmology have been identified, but not yet explored in detail. However, the examples of DM and GWs already indicate that AEDGE offers rich possibilities for scientific exploration and discovery.\\

The roadmap towards the AEDGE mission includes the following elements: 
\begin{itemize}

\item Today to 2025: Prototype 10-m facilities in the US, Europe and China, being extended to ${\cal O}(100)$m; 

\item  2025 to 2035: scaling of 100-m facilities to km-scale infrastructures;

\item  These experiments will demonstrate the reliability of cold-atom interferometers capable of achieving or surpassing the technical requirements for AEDGE;

\item  Operation of LISA will demonstrate the operation of large-scale laser interferometry in space;

\item In parallel, a vigorous technology development programme should be set up, pursued and coordinated on a European-wide level in order to maximize efficiency and avoid duplication. so as to build on the ground work laid by the development of ACES/PHARAO, the recent demonstration experiments of cold-atom and laser technology on rockets, and the laser technology development currently funded by ESA, and thereby continue the demonstrations by initial US, European and Chinese experiments of the  robustness of cold-atom technology in space.  \\
\end{itemize}

{\it AEDGE is a uniquely interdisciplinary mission that will harness cold atom technologies to address key issues in fundamental physics, astrophysics and cosmology that can be realized within the Voyage 2050 Science Programme of ESA. The worldwide spread of the authors of this article indicate that there could be global interest in participating in this mission.}

\newpage

\bibliographystyle{JHEP}
\bibliography{AEDGE}

\providecommand{\href}[2]{#2}\begingroup\raggedright\begin{thebibliography}{100}

\bibitem{workshop}
CERN, ``{ Workshop on Atomic Experiments for Dark Matter and Gravity
  Exploration}.'' \url{https://indico.cern.ch/event/830432/}.

\bibitem{Aghanim:2018eyx}
{\scshape Planck} collaboration, N.~Aghanim et~al., \emph{{Planck 2018 results.
  VI. Cosmological parameters}},
  \href{https://arxiv.org/abs/1807.06209}{{\ttfamily 1807.06209}}.

\bibitem{TheLIGOScientific:2014jea}
{\scshape LIGO Scientific} collaboration, J.~Aasi et~al., \emph{{Advanced
  LIGO}}, \href{https://doi.org/10.1088/0264-9381/32/7/074001}{\emph{Class.
  Quant. Grav.} {\bfseries 32} (2015) 074001},
  [\href{https://arxiv.org/abs/1411.4547}{{\ttfamily 1411.4547}}].

\bibitem{TheVirgo:2014hva}
{\scshape VIRGO} collaboration, F.~Acernese et~al., \emph{{Advanced Virgo: a
  second-generation interferometric gravitational wave detector}},
  \href{https://doi.org/10.1088/0264-9381/32/2/024001}{\emph{Class. Quant.
  Grav.} {\bfseries 32} (2015) 024001},
  [\href{https://arxiv.org/abs/1408.3978}{{\ttfamily 1408.3978}}].

\bibitem{Somiya:2011np}
{\scshape KAGRA} collaboration, K.~Somiya, \emph{{Detector configuration of
  KAGRA: The Japanese cryogenic gravitational-wave detector}},
  \href{https://doi.org/10.1088/0264-9381/29/12/124007}{\emph{Class. Quant.
  Grav.} {\bfseries 29} (2012) 124007},
  [\href{https://arxiv.org/abs/1111.7185}{{\ttfamily 1111.7185}}].

\bibitem{Unnikrishnan:2013qwa}
C.~S. Unnikrishnan, \emph{{IndIGO and LIGO-India: Scope and plans for
  gravitational wave research and precision metrology in India}},
  \href{https://doi.org/10.1142/S0218271813410101}{\emph{Int. J. Mod. Phys.}
  {\bfseries D22} (2013) 1341010},
  [\href{https://arxiv.org/abs/1510.06059}{{\ttfamily 1510.06059}}].

\bibitem{Punturo:2010zz}
M.~Punturo et~al., \emph{{The Einstein Telescope: A third-generation
  gravitational wave observatory}},
  \href{https://doi.org/10.1088/0264-9381/27/19/194002}{\emph{Class. Quant.
  Grav.} {\bfseries 27} (2010) 194002}.

\bibitem{Sathyaprakash:2012jk}
B.~Sathyaprakash et~al., \emph{{Scientific Objectives of Einstein Telescope}},
  \href{https://doi.org/10.1088/0264-9381/29/12/124013,
  10.1088/0264-9381/30/7/079501}{\emph{Class. Quant. Grav.} {\bfseries 29}
  (2012) 124013}, [\href{https://arxiv.org/abs/1206.0331}{{\ttfamily
  1206.0331}}].

\bibitem{Reitze:2019iox}
D.~Reitze et~al., \emph{{Cosmic Explorer: The U.S. Contribution to
  Gravitational-Wave Astronomy beyond LIGO}},
  \href{https://arxiv.org/abs/1907.04833}{{\ttfamily 1907.04833}}.

\bibitem{Guo:2018npi}
Z.-K. Guo, R.-G. Cai and Y.-Z. Zhang, \emph{{Taiji Program: Gravitational-Wave
  Sources}},  \href{https://arxiv.org/abs/1807.09495}{{\ttfamily 1807.09495}}.

\bibitem{Luo_2016}
J.~Luo, L.-S. Chen, H.-Z. Duan, Y.-G. Gong, S.~Hu, J.~Ji et~al.,
  \emph{{TianQin}: a space-borne gravitational wave detector},
  \href{https://doi.org/10.1088/0264-9381/33/3/035010}{\emph{Classical and
  Quantum Gravity} {\bfseries 33} (jan, 2016) 035010}.

\bibitem{Bender:2013nsa}
P.~L. Bender, M.~C. Begelman and J.~R. Gair, \emph{{Possible LISA follow-on
  mission scientific objectives}},
  \href{https://doi.org/10.1088/0264-9381/30/16/165017}{\emph{Class. Quant.
  Grav.} {\bfseries 30} (2013) 165017}.

\bibitem{Kawamura:2011zz}
S.~Kawamura et~al., \emph{{The Japanese space gravitational wave antenna:
  DECIGO}}, \href{https://doi.org/10.1088/0264-9381/28/9/094011}{\emph{Class.
  Quant. Grav.} {\bfseries 28} (2011) 094011}.

\bibitem{Mandel:2017pzd}
I.~Mandel, A.~Sesana and A.~Vecchio, \emph{{The astrophysical science case for
  a decihertz gravitational-wave detector}},
  \href{https://doi.org/10.1088/1361-6382/aaa7e0}{\emph{Class. Quant. Grav.}
  {\bfseries 35} (2018) 054004},
  [\href{https://arxiv.org/abs/1710.11187}{{\ttfamily 1710.11187}}].

\bibitem{Baker:2019pnp}
J.~Baker et~al., \emph{{Space Based Gravitational Wave Astronomy Beyond LISA
  (Astro2020 APC White Paper)}},
  \href{https://arxiv.org/abs/1907.11305}{{\ttfamily 1907.11305}}.

\bibitem{Kuns:2019upi}
K.~A. Kuns, H.~Yu, Y.~Chen and R.~X. Adhikari, \emph{{Astrophysics and
  cosmology with a deci-hertz gravitational-wave detector: TianGO}},
  \href{https://arxiv.org/abs/1908.06004}{{\ttfamily 1908.06004}}.

\bibitem{Pezze:2018wyn}
L.~Pezz{\` e}, A.~Smerzi, M.~K. Oberthaler, R.~Schmied and P.~Treutlein,
  \emph{{Quantum metrology with nonclassical states of atomic ensembles}},
  \href{https://doi.org/10.1103/RevModPhys.90.035005}{\emph{Rev. Mod. Phys.}
  {\bfseries 90} (2018) 035005},
  [\href{https://arxiv.org/abs/1609.01609}{{\ttfamily 1609.01609}}].

\bibitem{Aprile:2018dbl}
{\scshape XENON} collaboration, E.~Aprile et~al., \emph{{Dark Matter Search
  Results from a One Ton-Year Exposure of XENON1T}},
  \href{https://doi.org/10.1103/PhysRevLett.121.111302}{\emph{Phys. Rev. Lett.}
  {\bfseries 121} (2018) 111302},
  [\href{https://arxiv.org/abs/1805.12562}{{\ttfamily 1805.12562}}].

\bibitem{Battaglieri:2017aum}
M.~Battaglieri et~al., \emph{{US Cosmic Visions: New Ideas in Dark Matter 2017:
  Community Report}},  in \emph{{U.S. Cosmic Visions: New Ideas in Dark Matter
  College Park, MD, USA, March 23-25, 2017}}, 2017,
  \href{https://arxiv.org/abs/1707.04591}{{\ttfamily 1707.04591}},
  \href{http://lss.fnal.gov/archive/2017/conf/fermilab-conf-17-282-ae-ppd-t.pdf}{http://lss.fnal.gov/archive/2017/conf/fermilab-conf-17-282-ae-ppd-t.pdf}.

\bibitem{Preskill:1982cy}
J.~Preskill, M.~B. Wise and F.~Wilczek, \emph{{Cosmology of the Invisible
  Axion}}, \href{https://doi.org/10.1016/0370-2693(83)90637-8}{\emph{Phys.
  Lett.} {\bfseries B120} (1983) 127--132}.

\bibitem{Abbott:1982af}
L.~F. Abbott and P.~Sikivie, \emph{{A Cosmological Bound on the Invisible
  Axion}}, \href{https://doi.org/10.1016/0370-2693(83)90638-X}{\emph{Phys.
  Lett.} {\bfseries B120} (1983) 133--136}.

\bibitem{Dine:1982ah}
M.~Dine and W.~Fischler, \emph{{The Not So Harmless Axion}},
  \href{https://doi.org/10.1016/0370-2693(83)90639-1}{\emph{Phys. Lett.}
  {\bfseries B120} (1983) 137--141}.

\bibitem{Berge:2017ovy}
J.~Berge, P.~Brax, G.~M{\' e}tris, M.~Pernot-Borras, P.~Touboul and J.-P. Uzan,
  \emph{{MICROSCOPE Mission: First Constraints on the Violation of the Weak
  Equivalence Principle by a Light Scalar Dilaton}},
  \href{https://doi.org/10.1103/PhysRevLett.120.141101}{\emph{Phys. Rev. Lett.}
  {\bfseries 120} (2018) 141101},
  [\href{https://arxiv.org/abs/1712.00483}{{\ttfamily 1712.00483}}].

\bibitem{Hees:2018fpg}
A.~Hees, O.~Minazzoli, E.~Savalle, Y.~V. Stadnik and P.~Wolf, \emph{{Violation
  of the equivalence principle from light scalar dark matter}},
  \href{https://doi.org/10.1103/PhysRevD.98.064051}{\emph{Phys. Rev.}
  {\bfseries D98} (2018) 064051},
  [\href{https://arxiv.org/abs/1807.04512}{{\ttfamily 1807.04512}}].

\bibitem{Schlamminger:2007ht}
S.~Schlamminger, K.~Y. Choi, T.~A. Wagner, J.~H. Gundlach and E.~G. Adelberger,
  \emph{{Test of the equivalence principle using a rotating torsion balance}},
  \href{https://doi.org/10.1103/PhysRevLett.100.041101}{\emph{Phys. Rev. Lett.}
  {\bfseries 100} (2008) 041101},
  [\href{https://arxiv.org/abs/0712.0607}{{\ttfamily 0712.0607}}].

\bibitem{Wagner:2012ui}
T.~A. Wagner, S.~Schlamminger, J.~H. Gundlach and E.~G. Adelberger,
  \emph{{Torsion-balance tests of the weak equivalence principle}},
  \href{https://doi.org/10.1088/0264-9381/29/18/184002}{\emph{Class. Quant.
  Grav.} {\bfseries 29} (2012) 184002},
  [\href{https://arxiv.org/abs/1207.2442}{{\ttfamily 1207.2442}}].

\bibitem{VanTilburg:2015oza}
K.~Van~Tilburg, N.~Leefer, L.~Bougas and D.~Budker, \emph{{Search for
  ultralight scalar dark matter with atomic spectroscopy}},
  \href{https://doi.org/10.1103/PhysRevLett.115.011802}{\emph{Phys. Rev. Lett.}
  {\bfseries 115} (2015) 011802},
  [\href{https://arxiv.org/abs/1503.06886}{{\ttfamily 1503.06886}}].

\bibitem{Hees:2016gop}
A.~Hees, J.~Guena, M.~Abgrall, S.~Bize and P.~Wolf, \emph{{Searching for an
  oscillating massive scalar field as a dark matter candidate using atomic
  hyperfine frequency comparisons}},
  \href{https://doi.org/10.1103/PhysRevLett.117.061301}{\emph{Phys. Rev. Lett.}
  {\bfseries 117} (2016) 061301},
  [\href{https://arxiv.org/abs/1604.08514}{{\ttfamily 1604.08514}}].

\bibitem{Geraci:2016fva}
A.~A. Geraci and A.~Derevianko, \emph{{Sensitivity of atom interferometry to
  ultralight scalar field dark matter}},
  \href{https://doi.org/10.1103/PhysRevLett.117.261301}{\emph{Phys. Rev. Lett.}
  {\bfseries 117} (2016) 261301},
  [\href{https://arxiv.org/abs/1605.04048}{{\ttfamily 1605.04048}}].

\bibitem{Arvanitaki:2016fyj}
A.~Arvanitaki, P.~W. Graham, J.~M. Hogan, S.~Rajendran and K.~Van~Tilburg,
  \emph{{Search for light scalar dark matter with atomic gravitational wave
  detectors}}, \href{https://doi.org/10.1103/PhysRevD.97.075020}{\emph{Phys.
  Rev.} {\bfseries D97} (2018) 075020},
  [\href{https://arxiv.org/abs/1606.04541}{{\ttfamily 1606.04541}}].

\bibitem{Arvanitaki:2014faa}
A.~Arvanitaki, J.~Huang and K.~Van~Tilburg, \emph{{Searching for dilaton dark
  matter with atomic clocks}},
  \href{https://doi.org/10.1103/PhysRevD.91.015015}{\emph{Phys. Rev.}
  {\bfseries D91} (2015) 015015},
  [\href{https://arxiv.org/abs/1405.2925}{{\ttfamily 1405.2925}}].

\bibitem{Stadnik:2015kia}
Y.~V. Stadnik and V.~V. Flambaum, \emph{{Can dark matter induce cosmological
  evolution of the fundamental constants of Nature?}},
  \href{https://doi.org/10.1103/PhysRevLett.115.201301}{\emph{Phys. Rev. Lett.}
  {\bfseries 115} (2015) 201301},
  [\href{https://arxiv.org/abs/1503.08540}{{\ttfamily 1503.08540}}].

\bibitem{Damour:2010rm}
T.~Damour and J.~F. Donoghue, \emph{{Phenomenology of the Equivalence Principle
  with Light Scalars}},
  \href{https://doi.org/10.1088/0264-9381/27/20/202001}{\emph{Class. Quant.
  Grav.} {\bfseries 27} (2010) 202001},
  [\href{https://arxiv.org/abs/1007.2790}{{\ttfamily 1007.2790}}].

\bibitem{Damour:2010rp}
T.~Damour and J.~F. Donoghue, \emph{{Equivalence Principle Violations and
  Couplings of a Light Dilaton}},
  \href{https://doi.org/10.1103/PhysRevD.82.084033}{\emph{Phys. Rev.}
  {\bfseries D82} (2010) 084033},
  [\href{https://arxiv.org/abs/1007.2792}{{\ttfamily 1007.2792}}].

\bibitem{Stadnik:2014tta}
Y.~V. Stadnik and V.~V. Flambaum, \emph{{Searching for dark matter and
  variation of fundamental constants with laser and maser interferometry}},
  \href{https://doi.org/10.1103/PhysRevLett.114.161301}{\emph{Phys. Rev. Lett.}
  {\bfseries 114} (2015) 161301},
  [\href{https://arxiv.org/abs/1412.7801}{{\ttfamily 1412.7801}}].

\bibitem{Graham:2017ivz}
P.~W. Graham, D.~E. Kaplan, J.~Mardon, S.~Rajendran, W.~A. Terrano, L.~Trahms
  et~al., \emph{{Spin Precession Experiments for Light Axionic Dark Matter}},
  \href{https://doi.org/10.1103/PhysRevD.97.055006}{\emph{Phys. Rev.}
  {\bfseries D97} (2018) 055006},
  [\href{https://arxiv.org/abs/1709.07852}{{\ttfamily 1709.07852}}].

\bibitem{Graham:2015ifn}
P.~W. Graham, D.~E. Kaplan, J.~Mardon, S.~Rajendran and W.~A. Terrano,
  \emph{{Dark Matter Direct Detection with Accelerometers}},
  \href{https://doi.org/10.1103/PhysRevD.93.075029}{\emph{Phys. Rev.}
  {\bfseries D93} (2016) 075029},
  [\href{https://arxiv.org/abs/1512.06165}{{\ttfamily 1512.06165}}].

\bibitem{OHare:2018trr}
C.~A.~J. O'Hare, C.~McCabe, N.~W. Evans, G.~Myeong and V.~Belokurov,
  \emph{{Dark matter hurricane: Measuring the S1 stream with dark matter
  detectors}}, \href{https://doi.org/10.1103/PhysRevD.98.103006}{\emph{Phys.
  Rev.} {\bfseries D98} (2018) 103006},
  [\href{https://arxiv.org/abs/1807.09004}{{\ttfamily 1807.09004}}].

\bibitem{Roberts:2018agv}
B.~M. Roberts and A.~Derevianko, \emph{{Precision measurement noise asymmetry
  and its annual modulation as a dark matter signature}},
  \href{https://arxiv.org/abs/1803.00617}{{\ttfamily 1803.00617}}.

\bibitem{LIGOScientific:2018mvr}
{\scshape LIGO Scientific, Virgo} collaboration, B.~P. Abbott et~al.,
  \emph{{GWTC-1: A Gravitational-Wave Transient Catalog of Compact Binary
  Mergers Observed by LIGO and Virgo during the First and Second Observing
  Runs}},  \href{https://arxiv.org/abs/1811.12907}{{\ttfamily 1811.12907}}.

\bibitem{Audley:2017drz}
{\scshape LISA} collaboration, H.~Audley et~al., \emph{{Laser Interferometer
  Space Antenna}},  \href{https://arxiv.org/abs/1702.00786}{{\ttfamily
  1702.00786}}.

\bibitem{vanHaasteren:2011ni}
R.~van Haasteren et~al., \emph{{Placing limits on the stochastic
  gravitational-wave background using European Pulsar Timing Array data}},
  \href{https://doi.org/10.1111/j.1365-2966.2011.18613.x,
  10.1111/j.1365-2966.2012.20916.x}{\emph{Mon. Not. Roy. Astron. Soc.}
  {\bfseries 414} (2011) 3117--3128},
  [\href{https://arxiv.org/abs/1103.0576}{{\ttfamily 1103.0576}}].

\bibitem{Canuel:2017rrp}
B.~Canuel et~al., \emph{{Exploring gravity with the MIGA large scale atom
  interferometer}},
  \href{https://doi.org/10.1038/s41598-018-32165-z}{\emph{Sci. Rep.} {\bfseries
  8} (2018) 14064}, [\href{https://arxiv.org/abs/1703.02490}{{\ttfamily
  1703.02490}}].

\bibitem{Zhan:2019quq}
M.-S. Zhan et~al., \emph{{ZAIGA: Zhaoshan Long-baseline Atom Interferometer
  Gravitation Antenna}},
  \href{https://doi.org/10.1142/S0218271819400054}{\emph{Int. J. Mod. Phys.}
  {\bfseries D28} (2019) 1940005},
  [\href{https://arxiv.org/abs/1903.09288}{{\ttfamily 1903.09288}}].

\bibitem{Graham:2017pmn}
{\scshape MAGIS} collaboration, P.~W. Graham, J.~M. Hogan, M.~A. Kasevich,
  S.~Rajendran and R.~W. Romani, \emph{{Mid-band gravitational wave detection
  with precision atomic sensors}},
  \href{https://arxiv.org/abs/1711.02225}{{\ttfamily 1711.02225}}.

\bibitem{Bouyer}
P.~Bouyer, ``{MIGA and ELGAR: New Perspectives for Low Frequency Gravitational
  Wave Observation Using Atom Interferometry}.''
  \url{https://indico.obspm.fr/event/58/contributions/214/attachments/88/98/Slides-bouyer2018\_06\_21\_MIGA\_GDR.pdf},
  2018.

\bibitem{AION}
{\scshape AION Core Team} collaboration, K.~Bongs et~al., ``{An Atom
  Interferometer Observatory and Network (AION) for the exploration of
  Ultra-Light Dark Matter and Mid-Frequency Gravitational Waves}.''
  \url{https://www.hep.ph.ic.ac.uk/AION-Project/}, 2019.

\bibitem{Magorrian:1997hw}
J.~Magorrian et~al., \emph{{The Demography of massive dark objects in galaxy
  centers}}, \href{https://doi.org/10.1086/300353}{\emph{Astron. J.} {\bfseries
  115} (1998) 2285}, [\href{https://arxiv.org/abs/astro-ph/9708072}{{\ttfamily
  astro-ph/9708072}}].

\bibitem{Kauffmann:1999ce}
G.~Kauffmann and M.~Haehnelt, \emph{{A Unified model for the evolution of
  galaxies and quasars}},
  \href{https://doi.org/10.1046/j.1365-8711.2000.03077.x}{\emph{Mon. Not. Roy.
  Astron. Soc.} {\bfseries 311} (2000) 576--588},
  [\href{https://arxiv.org/abs/astro-ph/9906493}{{\ttfamily
  astro-ph/9906493}}].

\bibitem{Akiyama:2019cqa}
{\scshape Event Horizon Telescope} collaboration, K.~Akiyama et~al.,
  \emph{{First M87 Event Horizon Telescope Results. I. The Shadow of the
  Supermassive Black Hole}},
  \href{https://doi.org/10.3847/2041-8213/ab0ec7}{\emph{Astrophys. J.}
  {\bfseries 875} (2019) L1}.

\bibitem{Rees:1984si}
M.~J. Rees, \emph{{Black Hole Models for Active Galactic Nuclei}},
  \href{https://doi.org/10.1146/annurev.aa.22.090184.002351}{\emph{Ann. Rev.
  Astron. Astrophys.} {\bfseries 22} (1984) 471--506}.

\bibitem{Mezcua:2017npy}
M.~Mezcua, \emph{{Observational evidence for intermediate-mass black holes}},
  \href{https://doi.org/10.1142/S021827181730021X}{\emph{Int. J. Mod. Phys.}
  {\bfseries D26} (2017) 1730021},
  [\href{https://arxiv.org/abs/1705.09667}{{\ttfamily 1705.09667}}].

\bibitem{KSH}
D.~Katz, H.~Sijacki and M.~Haehnelt, \emph{{Seeding High Redshift QSOs by
  Collisional Runaway in Primordial Star Clusters}}, {\emph{MNRAS} {\bfseries
  451} (2015) 2352}.

\bibitem{Volonteri:2002vz}
M.~Volonteri, F.~Haardt and P.~Madau, \emph{{The Assembly and merging history
  of supermassive black holes in hierarchical models of galaxy formation}},
  \href{https://doi.org/10.1086/344675}{\emph{Astrophys. J.} {\bfseries 582}
  (2003) 559--573}, [\href{https://arxiv.org/abs/astro-ph/0207276}{{\ttfamily
  astro-ph/0207276}}].

\bibitem{Volonteri:2007ax}
M.~Volonteri, G.~Lodato and P.~Natarajan, \emph{{The evolution of massive black
  hole seeds}},
  \href{https://doi.org/10.1111/j.1365-2966.2007.12589.x}{\emph{Mon. Not. Roy.
  Astron. Soc.} {\bfseries 383} (2008) 1079},
  [\href{https://arxiv.org/abs/0709.0529}{{\ttfamily 0709.0529}}].

\bibitem{Junca:2019xvb}
J.~Junca et~al., \emph{{Characterizing Earth gravity field fluctuations with
  the MIGA antenna for future Gravitational Wave detectors}},
  \href{https://doi.org/10.1103/PhysRevD.99.104026}{\emph{Phys. Rev.}
  {\bfseries D99} (2019) 104026},
  [\href{https://arxiv.org/abs/1902.05337}{{\ttfamily 1902.05337}}].

\bibitem{Chaibi:2016dze}
W.~Chaibi, R.~Geiger, B.~Canuel, A.~Bertoldi, A.~Landragin and P.~Bouyer,
  \emph{{Low Frequency Gravitational Wave Detection With Ground Based Atom
  Interferometer Arrays}},
  \href{https://doi.org/10.1103/PhysRevD.93.021101}{\emph{Phys. Rev.}
  {\bfseries D93} (2016) 021101},
  [\href{https://arxiv.org/abs/1601.00417}{{\ttfamily 1601.00417}}].

\bibitem{Erickcek:2006xc}
A.~L. Erickcek, M.~Kamionkowski and A.~J. Benson, \emph{{Supermassive Black
  Hole Merger Rates: Uncertainties from Halo Merger Theory}},
  \href{https://doi.org/10.1111/j.1365-2966.2006.10838.x}{\emph{Mon. Not. Roy.
  Astron. Soc.} {\bfseries 371} (2006) 1992--2000},
  [\href{https://arxiv.org/abs/astro-ph/0604281}{{\ttfamily
  astro-ph/0604281}}].

\bibitem{Heger:2002by}
A.~Heger, C.~L. Fryer, S.~E. Woosley, N.~Langer and D.~H. Hartmann, \emph{{How
  massive single stars end their life}},
  \href{https://doi.org/10.1086/375341}{\emph{Astrophys. J.} {\bfseries 591}
  (2003) 288--300}, [\href{https://arxiv.org/abs/astro-ph/0212469}{{\ttfamily
  astro-ph/0212469}}].

\bibitem{Sesana2016}
A.~Sesana, \emph{Prospects for multiband gravitational-wave astronomy after
  {GW}150914},
  \href{https://doi.org/10.1103/physrevlett.116.231102}{\emph{Physical Review
  Letters} {\bfseries 116} (June, 2016) }.

\bibitem{Graham:2017lmg}
P.~W. Graham and S.~Jung, \emph{{Localizing Gravitational Wave Sources with
  Single-Baseline Atom Interferometers}},
  \href{https://doi.org/10.1103/PhysRevD.97.024052}{\emph{Phys. Rev.}
  {\bfseries D97} (2018) 024052},
  [\href{https://arxiv.org/abs/1710.03269}{{\ttfamily 1710.03269}}].

\bibitem{Carson:2019rda}
Z.~Carson and K.~Yagi, \emph{{Multi-band gravitational wave tests of general
  relativity}},  \href{https://arxiv.org/abs/1905.13155}{{\ttfamily
  1905.13155}}.

\bibitem{Bern:2019nnu}
Z.~Bern, C.~Cheung, R.~Roiban, C.-H. Shen, M.~P. Solon and M.~Zeng,
  \emph{{Scattering Amplitudes and the Conservative Hamiltonian for Binary
  Systems at Third Post-Minkowskian Order}},
  \href{https://doi.org/10.1103/PhysRevLett.122.201603}{\emph{Phys. Rev. Lett.}
  {\bfseries 122} (2019) 201603},
  [\href{https://arxiv.org/abs/1901.04424}{{\ttfamily 1901.04424}}].

\bibitem{Burdge:2019hgl}
K.~B. Burdge et~al., \emph{{General relativistic orbital decay in a
  seven-minute-orbital-period eclipsing binary system}},
  \href{https://doi.org/10.1038/s41586-019-1403-0}{\emph{Nature} {\bfseries
  571} (2019) 528--531}, [\href{https://arxiv.org/abs/1907.11291}{{\ttfamily
  1907.11291}}].

\bibitem{Ellis:2019tjf}
J.~Ellis, M.~Fairbairn, M.~Lewicki, V.~Vaskonen and A.~Wickens,
  \emph{{Intergalactic Magnetic Fields from First-Order Phase Transitions}},
  \href{https://arxiv.org/abs/1907.04315}{{\ttfamily 1907.04315}}.

\bibitem{Ellis:2018mja}
J.~Ellis, M.~Lewicki and J.~M. No, \emph{{On the Maximal Strength of a
  First-Order Electroweak Phase Transition and its Gravitational Wave Signal}},
   \href{https://arxiv.org/abs/1809.08242}{{\ttfamily 1809.08242}}.

\bibitem{Ellis:2019oqb}
J.~Ellis, M.~Lewicki, J.~M. No and V.~Vaskonen, \emph{{Gravitational wave
  energy budget in strongly supercooled phase transitions}},
  \href{https://arxiv.org/abs/1903.09642}{{\ttfamily 1903.09642}}.

\bibitem{Abada:2019lih}
{\scshape FCC} collaboration, A.~Abada et~al., \emph{{FCC Physics
  Opportunities}},
  \href{https://doi.org/10.1140/epjc/s10052-019-6904-3}{\emph{Eur. Phys. J.}
  {\bfseries C79} (2019) 474}.

\bibitem{Bacon:2018dui}
{\scshape SKA} collaboration, D.~J. Bacon et~al., \emph{{Cosmology with Phase 1
  of the Square Kilometre Array: Red Book 2018: Technical specifications and
  performance forecasts}}, {\emph{Submitted to: Publ. Astron. Soc. Austral.}
  (2018) }, [\href{https://arxiv.org/abs/1811.02743}{{\ttfamily 1811.02743}}].

\bibitem{Cui:2018rwi}
Y.~Cui, M.~Lewicki, D.~E. Morrissey and J.~D. Wells, \emph{{Probing the pre-BBN
  universe with gravitational waves from cosmic strings}},
  \href{https://doi.org/10.1007/JHEP01(2019)081}{\emph{JHEP} {\bfseries 01}
  (2019) 081}, [\href{https://arxiv.org/abs/1808.08968}{{\ttfamily
  1808.08968}}].

\bibitem{Tino2019}
G.~M. Tino et~al., \emph{{SAGE: A Proposal for a Space Atomic Gravity
  Explorer}}, {\emph{Eur. Phys. J. D} {\bfseries Topical Issue on Quantum
  Technologies for Gravitational Physics} (2019) In press},
  [\href{https://arxiv.org/abs/1907.03867}{{\ttfamily 1907.03867}}].

\bibitem{Becker:2007sv}
J.~K. Becker, \emph{{High-energy neutrinos in the context of multimessenger
  physics}}, \href{https://doi.org/10.1016/j.physrep.2007.10.006}{\emph{Phys.
  Rept.} {\bfseries 458} (2008) 173--246},
  [\href{https://arxiv.org/abs/0710.1557}{{\ttfamily 0710.1557}}].

\bibitem{Alonso:2018dxy}
R.~Alonso, D.~Blas and P.~Wolf, \emph{{Exploring the ultra-light to sub-MeV
  dark matter window with atomic clocks and co-magnetometers}},
  \href{https://arxiv.org/abs/1810.00889}{{\ttfamily 1810.00889}}.

\bibitem{Peters_2001}
A.~Peters, K.~Y. Chung and S.~Chu, \emph{High-precision gravity measurements
  using atom interferometry},
  \href{https://doi.org/10.1088/0026-1394/38/1/4}{\emph{Metrologia} {\bfseries
  38} (feb, 2001) 25--61}.

\bibitem{Safronova:2017xyt}
M.~S. Safronova, D.~Budker, D.~DeMille, D.~F.~J. Kimball, A.~Derevianko and
  C.~W. Clark, \emph{{Search for New Physics with Atoms and Molecules}},
  \href{https://doi.org/10.1103/RevModPhys.90.025008}{\emph{Rev. Mod. Phys.}
  {\bfseries 90} (2018) 025008},
  [\href{https://arxiv.org/abs/1710.01833}{{\ttfamily 1710.01833}}].

\bibitem{Dimopoulos:2008hx}
S.~Dimopoulos, P.~W. Graham, J.~M. Hogan and M.~A. Kasevich, \emph{{General
  Relativistic Effects in Atom Interferometry}},
  \href{https://doi.org/10.1103/PhysRevD.78.042003}{\emph{Phys. Rev.}
  {\bfseries D78} (2008) 042003},
  [\href{https://arxiv.org/abs/0802.4098}{{\ttfamily 0802.4098}}].

\bibitem{Uzan:2002vq}
J.-P. Uzan, \emph{{The Fundamental constants and their variation: Observational
  status and theoretical motivations}},
  \href{https://doi.org/10.1103/RevModPhys.75.403}{\emph{Rev. Mod. Phys.}
  {\bfseries 75} (2003) 403},
  [\href{https://arxiv.org/abs/hep-ph/0205340}{{\ttfamily hep-ph/0205340}}].

\bibitem{Martins:2019qxe}
C.~J. A.~P. Martins and M.~V. Miñana, \emph{{Consistency of local and
  astrophysical tests of the stability of fundamental constants}},
  \href{https://doi.org/10.1016/j.dark.2019.100301}{\emph{Phys. Dark Univ.}
  {\bfseries 25} (2019) 100301},
  [\href{https://arxiv.org/abs/1904.07896}{{\ttfamily 1904.07896}}].

\bibitem{Jaffe:2016fsh}
M.~Jaffe, P.~Haslinger, V.~Xu, P.~Hamilton, A.~Upadhye, B.~Elder et~al.,
  \emph{{Testing sub-gravitational forces on atoms from a miniature, in-vacuum
  source mass}}, \href{https://doi.org/10.1038/nphys4189}{\emph{Nature Phys.}
  {\bfseries 13} (2017) 938},
  [\href{https://arxiv.org/abs/1612.05171}{{\ttfamily 1612.05171}}].

\bibitem{Sabulsky:2018jma}
D.~O. Sabulsky, I.~Dutta, E.~A. Hinds, B.~Elder, C.~Burrage and E.~J. Copeland,
  \emph{{Experiment to detect dark energy forces using atom interferometry}},
  \href{https://arxiv.org/abs/1812.08244}{{\ttfamily 1812.08244}}.

\bibitem{Ellis:1983jz}
J.~R. Ellis, J.~S. Hagelin, D.~V. Nanopoulos and M.~Srednicki, \emph{{Search
  for Violations of Quantum Mechanics}},
  \href{https://doi.org/10.1016/0550-3213(84)90053-1}{\emph{Nucl. Phys.}
  {\bfseries B241} (1984) 381}.

\bibitem{Banks:1983by}
T.~Banks, L.~Susskind and M.~E. Peskin, \emph{{Difficulties for the Evolution
  of Pure States Into Mixed States}},
  \href{https://doi.org/10.1016/0550-3213(84)90184-6}{\emph{Nucl. Phys.}
  {\bfseries B244} (1984) 125--134}.

\bibitem{Ghirardi:1985mt}
G.~C. Ghirardi, A.~Rimini and T.~Weber, \emph{{A Unified Dynamics for Micro and
  MACRO Systems}}, \href{https://doi.org/10.1103/PhysRevD.34.470}{\emph{Phys.
  Rev.} {\bfseries D34} (1986) 470}.

\bibitem{Weinberg:2016uml}
S.~Weinberg, \emph{{Lindblad Decoherence in Atomic Clocks}},
  \href{https://doi.org/10.1103/PhysRevA.94.042117}{\emph{Phys. Rev.}
  {\bfseries A94} (2016) 042117},
  [\href{https://arxiv.org/abs/1610.02537}{{\ttfamily 1610.02537}}].

\bibitem{Chung:2009rm}
K.-Y. Chung, S.-w. Chiow, S.~Herrmann, S.~Chu and H.~Muller, \emph{{Atom
  interferometry tests of local Lorentz invariance in gravity and
  electrodynamics}},
  \href{https://doi.org/10.1103/PhysRevD.80.016002}{\emph{Phys. Rev.}
  {\bfseries D80} (2009) 016002},
  [\href{https://arxiv.org/abs/0905.1929}{{\ttfamily 0905.1929}}].

\bibitem{Graham:2012sy}
P.~W. Graham, J.~M. Hogan, M.~A. Kasevich and S.~Rajendran, \emph{{A New Method
  for Gravitational Wave Detection with Atomic Sensors}},
  \href{https://doi.org/10.1103/PhysRevLett.110.171102}{\emph{Phys. Rev. Lett.}
  {\bfseries 110} (2013) 171102},
  [\href{https://arxiv.org/abs/1206.0818}{{\ttfamily 1206.0818}}].

\bibitem{Graham:2016plp}
P.~W. Graham, J.~M. Hogan, M.~A. Kasevich and S.~Rajendran, \emph{{Resonant
  mode for gravitational wave detectors based on atom interferometry}},
  \href{https://doi.org/10.1103/PhysRevD.94.104022}{\emph{Phys. Rev.}
  {\bfseries D94} (2016) 104022},
  [\href{https://arxiv.org/abs/1606.01860}{{\ttfamily 1606.01860}}].

\bibitem{Hu:2017kp}
L.~Hu, N.~Poli, L.~Salvi and G.~M. Tino, \emph{{Atom Interferometry with the Sr
  Optical Clock Transition}}, {\emph{Physical Review Letters} {\bfseries 119}
  (Dec., 2017) 555--5}.

\bibitem{Snadden:1998zz}
M.~J. Snadden, J.~M. McGuirk, P.~Bouyer, K.~G. Haritos and M.~A. Kasevich,
  \emph{{Measurement of the Earth's Gravity Gradient with an Atom
  Interferometer-Based Gravity Gradiometer}},
  \href{https://doi.org/10.1103/PhysRevLett.81.971}{\emph{Phys. Rev. Lett.}
  {\bfseries 81} (1998) 971--974}.

\bibitem{Sorrentino:2013uza}
F.~Sorrentino, Q.~Bodart, L.~Cacciapuoti, Y.~H. Lien, M.~Prevedelli, G.~Rosi
  et~al., \emph{{Sensitivity limits of a Raman atom interferometer as a gravity
  gradiometer}}, \href{https://doi.org/10.1103/PhysRevA.89.023607}{\emph{Phys.
  Rev.} {\bfseries A89} (2014) 023607},
  [\href{https://arxiv.org/abs/1312.3741}{{\ttfamily 1312.3741}}].

\bibitem{Hogan:2015xla}
J.~M. Hogan and M.~A. Kasevich, \emph{{Atom interferometric gravitational wave
  detection using heterodyne laser links}},
  \href{https://doi.org/10.1103/PhysRevA.94.033632}{\emph{Phys. Rev.}
  {\bfseries A94} (2016) 033632},
  [\href{https://arxiv.org/abs/1501.06797}{{\ttfamily 1501.06797}}].

\bibitem{Dimopoulos:2008sv}
S.~Dimopoulos, P.~W. Graham, J.~M. Hogan, M.~A. Kasevich and S.~Rajendran,
  \emph{{An Atomic Gravitational Wave Interferometric Sensor (AGIS)}},
  \href{https://doi.org/10.1103/PhysRevD.78.122002}{\emph{Phys. Rev.}
  {\bfseries D78} (2008) 122002},
  [\href{https://arxiv.org/abs/0806.2125}{{\ttfamily 0806.2125}}].

\bibitem{Thrane:2013oya}
E.~Thrane and J.~D. Romano, \emph{{Sensitivity curves for searches for
  gravitational-wave backgrounds}},
  \href{https://doi.org/10.1103/PhysRevD.88.124032}{\emph{Phys. Rev.}
  {\bfseries D88} (2013) 124032},
  [\href{https://arxiv.org/abs/1310.5300}{{\ttfamily 1310.5300}}].

\bibitem{Liu:2018kn}
L.~Liu, D.-S. L{\"u}, W.-B. Chen, T.~Li, Q.-Z. Qu, B.~Wang et~al.,
  \emph{{In-orbit operation of an atomic clock based on laser-cooled 87Rb
  atoms}}, {\emph{Nature Communications} (July, 2018) 1--8}.

\bibitem{MAIUS}
D.~Becker et~al., \emph{{Space-borne Bose–Einstein Condensation for Precision
  Interferometry}}, {\emph{Nature} {\bfseries 562} (2018) 391}.

\bibitem{CAL}
E.~R. Elliott, M.~C. Krutzik, J.~R. Williams, R.~J. Thompson and D.~C. Aveline,
  \emph{Nasa's cold atom lab (cal): system development and ground test status},
  {\emph{npj Microgravity} {\bfseries 4} (Aug, 2016) 16}.

\bibitem{Lezius}
M.~Lezius et~al., \emph{Space-borne frequency comb metrology},
  \href{https://doi.org/10.1038/s41526-018-0049-9}{\emph{Optica} {\bfseries 3}
  (Dec, 2016) 1381}.

\bibitem{Dinkelaker}
A.~Dinkelaker et~al., \emph{Space-borne frequency comb metrology},
  \href{https://doi.org/10.1364/AO.56.001388}{\emph{Appl. Opt.} {\bfseries 56}
  (Feb, 2017) 1388}.

\bibitem{PhysRevApplied.11.054068}
K.~D\"oringshoff et~al., \emph{Iodine frequency reference on a sounding
  rocket}, \href{https://doi.org/10.1103/PhysRevApplied.11.054068}{\emph{Phys.
  Rev. Applied} {\bfseries 11} (May, 2019) 054068}.

\bibitem{Rosi:2014kva}
G.~Rosi, F.~Sorrentino, L.~Cacciapuoti, M.~Prevedelli and G.~M. Tino,
  \emph{{Precision Measurement of the Newtonian Gravitational Constant Using
  Cold Atoms}}, \href{https://doi.org/10.1038/nature13433}{\emph{Nature}
  {\bfseries 510} (2014) 518},
  [\href{https://arxiv.org/abs/1412.7954}{{\ttfamily 1412.7954}}].

\bibitem{MAGIA}
G.~T. Tino et~al., ``{Ultracold atoms and precision measurements}.''
  \url{http://coldatoms.lens.unifi.it}.

\bibitem{Sabulsky2}
{\scshape ELGAR} collaboration, D.~Sabulsky, ``{The European Laboratory for
  Gravitation and Atom- interferometric Research (ELGAR) Project}.''
  \url{https://indico.cern.ch/event/830432/contributions/3497166/attachments/1883894/3104651/sabulsky_ELGAR_CERN_2019.pdf},
  2019.

\bibitem{Balas}
R.~Kishor~Kumar et~al., \emph{{C and Fortran OpenMP programs for rotating
  Bose-Einstein condensates}},
  \href{https://arxiv.org/abs/1908.06327}{{\ttfamily 1908.06327}}.

\bibitem{ACES}
L.~Cacciapuoti and C.~Salomon, \emph{{The ACES experiment}}, {\emph{European
  Physical Journal - Special Topics} {\bfseries 172} (2009) 57}.

\bibitem{ACES-PHARAO}
P.~Laurent, D.~Massonnet, L.~Cacciapuoti and C.~Salomon, \emph{{The ACES/PHARAO
  space mission}}, {\emph{Comptes Rendus Physique} {\bfseries 16 (5)} (2015)
  540}.

\bibitem{Muntinga:2013pta}
H.~M{\" u}ntinga et~al., \emph{{Interferometry with Bose-Einstein Condensates
  in Microgravity}},
  \href{https://doi.org/10.1103/PhysRevLett.110.093602}{\emph{Phys. Rev. Lett.}
  {\bfseries 110} (2013) 093602},
  [\href{https://arxiv.org/abs/1301.5883}{{\ttfamily 1301.5883}}].

\bibitem{BECCAL}
{\scshape BECCAL} collaboration, D.~Becker et~al., ``{BECCAL Science
  Overview}.''
  \url{https://custom.cvent.com/216E523D934443CA9F514B796474A210/files/f7a0cce2d06f4e2182eaec7af912d5bf.pdf},
  2019.

\bibitem{Barrett2016}
B.~Barrett, L.~Antoni-Micollier, L.~Chichet, B.~Battelier,
  T.~L{\'{e}}v{\`{e}}que, A.~Landragin et~al., \emph{Dual matter-wave inertial
  sensors in weightlessness},
  \href{https://doi.org/10.1038/ncomms13786}{\emph{Nature Communications}
  {\bfseries 7} (Dec., 2016) }.

\bibitem{Condon2019}
G.~Condon, M.~Rabault, B.~Barrett, L.~Chichet, R.~Arguel, H.~Eneriz-Imaz
  et~al., \emph{{All-Optical Bose-Einstein Condensates in Microgravity}},
  \href{https://arxiv.org/abs/1906.10063 [physics.atom-ph]}{{\ttfamily
  1906.10063 [physics.atom-ph]}}.

\bibitem{SOC1}
K.~Bongs et~al., \emph{{Development of a strontium optical lattice clock for
  the SOC mission on the ISS}}, {\emph{Comptes Rendus Physique} {\bfseries 16
  (5)} (2015) 553}.

\bibitem{SOC2}
S.~Origlia et~al., \emph{{Development of a strontium optical lattice clock for
  the SOC mission on the ISS}}, {\emph{Phys. Rev. A} {\bfseries 98} (2015)
  053443}.

\bibitem{STE-QUEST}
D.~Aguilera et~al., \emph{{STE-QUEST - Test of the Universality of Free Fall
  Using Cold Atom Interferometry}},
  \href{https://doi.org/10.1088/0264-9381/31/11/115010,
  10.1088/0264-9381/31/15/159502}{\emph{Class. Quant. Grav.} {\bfseries 31}
  (2014) 115010}, [\href{https://arxiv.org/abs/1312.5980}{{\ttfamily
  1312.5980}}].

\bibitem{Lacour:2018nws}
S.~Lacour et~al., \emph{{SAGE: finding IMBH in the black hole desert}},
  \href{https://arxiv.org/abs/1811.04743}{{\ttfamily 1811.04743}}.

\bibitem{AIGSO}
D.~Gao, J.~Wang and M.~Zhan, \emph{{Atomic Interferometric Gravitational-Wave
  Space Observatory (AIGSO)}}, {\emph{Commun. Theor. Phys.} {\bfseries 69}
  (2018) 37}.

\bibitem{SAGAS}
P.~Wolf et~al., \emph{{Quantum physics exploring gravity in the outer solar
  system: the SAGAS project}}, {\emph{Experimental Astronomy} {\bfseries 23(2)}
  (2009) 651}.

\bibitem{SAI1}
G.~M. Tino et~al., \emph{{Atom interferometers and optical atomic clocks: New
  quantum sensors for fundamental physics experiments in space}},
  {\emph{Nuclear Physics B - Proceedings Supplements} {\bfseries 166} (2007)
  159}.

\bibitem{SAI2}
F.~Sorrentino et~al., \emph{{A Compact Atom Interferometer for Future Space
  Missions}}, {\emph{Microgravity Science and Technology} {\bfseries 22} (2010)
  551}.

\end{thebibliography}\endgroup



\providecommand{\href}[2]{#2}\begingroup\raggedright\endgroup

\end{document}